\documentclass[fleqn,usenatbib]{mnras}

\makeatletter
\renewcommand*{\@fnsymbol}[1]{\ensuremath{\ifcase#1\or *\or \dagger\or \ddagger\or
   \mathsection\or \mathparagraph\or \|\or **\or \dagger\dagger
   \or \ddagger\ddagger \else\@ctrerr\fi}}
\makeatother

\usepackage{graphicx}
\usepackage{ifthen}
\usepackage{url}
\usepackage{amsmath}
\usepackage{bm}
\usepackage{lscape}
\usepackage{rotating}
\usepackage[graphicx]{realboxes}
\usepackage{supertabular}
\usepackage{pdfpages}
\usepackage{footnote}
\usepackage{hyperref}
\usepackage{amssymb}
\makesavenoteenv{tabular}
\makesavenoteenv{table}
\usepackage[T1]{fontenc}
\usepackage{tipa}
\usepackage{tablefootnote}

\hypersetup{
  colorlinks   = true,    % Colours links instead of ugly boxes
  urlcolor     = blue,    % Colour for external hyperlinks
  linkcolor    = blue,    % Colour of internal links
  citecolor    = blue      % Colour of citations
}

\usepackage{ulem}

\def\ltsima{$\; \buildrel < \over \sim \;$}
\def\lta{\lower.5ex\hbox{\ltsima}}
\def\gtsima{$\; \buildrel > \over \sim \;$}
\def\simgt{\lower.5ex\hbox{\gtsima}}

\def\kms{{\rm\,km \  s^{-1}}}

\def\kpc{{\rm\,kpc}}

\def\msun{{\rm\,M_\odot}}

\def\teff{{\rm\,T_{eff}}}

\def\eg{{e.g.,\ }}
\def\ie{{i.e.,\ }}
\def\aka{{a.k.a.\ }}

\usepackage{amsmath}

\def\FeH{{\rm[Fe/H]}}

\def\ione{{~\sc i}}
\def\ii{{~\sc ii}}

\title[An ancient system hidden in the Galactic plane?]{An ancient system hidden in the Galactic plane?} 
\author[F. Sestito et al.] {Federico Sestito$^{1,2}$\thanks{Email: \url{f.sestito@herts.ac.uk}},
Emma Fern\'andez-Alvar$^{3,4}$,
Rebecca Brooks$^{2}$,
Emma Olson$^{2}$,
\newauthor
Leticia Carigi$^{5}$,
Paula Jofr\'e$^{6}$,
Danielle de Brito Silva$^{7,6}$,
Camilla J. L. Eldridge$^{6}$,
\newauthor
Sara Vitali$^{8,6}$,
Kim A. Venn$^{2}$,
Vanessa Hill$^{9}$,
Anke Ardern-Arentsen$^{10}$, 
\newauthor
Georges Kordopatis$^{9}$,
Nicolas F. Martin$^{11,12}$,
Julio F. Navarro$^{2}$,
Else Starkenburg$^{13}$,
\newauthor
Patricia B. Tissera$^{14,15}$,
Pascale Jablonka$^{16,17}$,
Carmela Lardo$^{18}$,
Romain Lucchesi$^{19}$,
\newauthor
Tobias Buck$^{20,21}$, and
Alexia Amayo$^{5}$\\
(Affiliations can be found before the references)
}

\pubyear{2026}
\date{Accepted XXX. Received YYY; in original form ZZZ}
\begin{document}

\label{firstpage}
\pagerange{\pageref{firstpage}--\pageref{lastpage}}

\maketitle

\begin{abstract}
We analyse high signal-to-noise ESPaDOnS/CFHT spectra of 20 very metal-poor stars (VMP; [Fe/H]~$<-2.0$) in the solar neighbourhood (within $\sim2$ kpc), selected to be on planar orbits with maximum heights $\lesssim4$ kpc. The sample comprises 11 stars on prograde and 9 on retrograde orbits, all with relatively high eccentricities (0.5--0.9).Their chemical abundance patterns indicate enrichment from high-energy supernovae and hypernovae up to the Fe-peak, and contributions from fast-rotating massive stars and neutron star mergers for the neutron-capture elements. No significant chemical differences are found between prograde and retrograde stars. 
The [Sr, Ba, Eu/Fe] ratios resemble those of stars in classical dwarfs galaxies.
Chemical dispersion and distance analyses further highlight the internal similarity of the sample and its separation from the bulk of the observed, non-planar halo population. Applying the same kinematical selection to another homogeneous dataset yields consistent results, confirming that this group of planar VMP stars exhibit peculiar chemical properties distinct from those of the observed halo and other known Galactic structures. 
These findings suggest that the stars formed in an environment that experienced a homogeneous chemical evolution akin to that of dwarf galaxies. 
A plausible scenario, supported by cosmological zoom-in simulations, is the early accretion of a single system whose subsequent dynamical evolution naturally produced stars on both prograde and retrograde planar orbits. If this interpretation is correct, we tentatively refer to this putative progenitor as \textit{Loki}. However, comparisons with other planar VMP stars spanning a wider range of chemo-dynamical properties indicate that multiple accretion events likely contributed to this diverse population orbiting close to the Galactic plane. 
\end{abstract}

\begin{keywords}
Galaxy: formation - Galaxy: evolution - stars: abundances - stars: kinematics and dynamics - stars: Population III - stars: Population II
\end{keywords}

\section{Introduction}\label{sec:introduction}
A way to understand the Galactic chemical and assembly history is to study the relics from those early epochs; these are assumed to be the most metal-poor stars \citep{Freeman02,Tolstoy09,Tumlinson10,Wise12}. Cosmological simulations suggest that the most metal-poor stars formed within $2-3$~Gyr from the Big Bang in low-mass systems, often called the building blocks of Milky Way-like galaxies \citep[\eg][]{Starkenburg17a,ElBadry18,Sestito21}. These building blocks merged together at early epochs, dispersing their stellar, gaseous, and dark matter content into the forming proto-Galaxy. Therefore, the most metal-poor stars coming from the early Galactic assembly are supposed to populate the inner regions of the Milky Way, while those accreted later might be dispersed in the outer halo \citep[][]{Bullock2005,Tissera12,Starkenburg17a}. 

These expectations are consistent with observations from various all-sky surveys, \eg SDSS \citep{York00} and LAMOST \citep{Cui12}, and from dedicated metal-poor surveys, \eg Pristine \citep{Starkenburg17b,Martin24} and SkyMapper \citep{DaCosta19}. Some of the most metal-poor stars have been serendipitously discovered in all-sky surveys \citep[\eg][]{Caffau11,Li15}, although they are extremely rare objects \citep{Youakim17} hidden in a haystack of more metal-rich and younger stars. To weed out the contamination, dedicated  metal-poor star-finding surveys use a metallicity-sensitive filter that includes the Ca HK spectral lines which, in combination with other broad-band filters, can be used to estimate a photometric \FeH\footnote{\FeH $= \log(\rm{N_{Fe}/N_{H}})_{\star}-\log(\rm{N_{Fe}/N_{H}})_{\odot} $, in which $\rm{N_X}$ is the number density of element X.}.  SkyMapper uses a  v-filter, while Pristine has a narrow-band filter centred on the Ca HK lines. In both cases, the two surveys reach a very high-efficiency in discovering the most metal-poor stars \citep{DaCosta19,Aguado19}. More recently, and with strategies similar to those based on narrow filters, J-Plus \citep{Cenarro19} and S-Plus \citep{Almeida-Fernandes22} are discovering and investigating new low-metallicity stars in the Milky Way \citep[\eg][]{Galarza22,Placco22,Perottoni24}.

A long-debated question in Galactic studies is whether the MW disc extends to the very metal-poor regimes \citep[VMP, $\FeH\leq-2.0$,][]{Bonifacio99,Beers02,Kordopatis13}. The synergy of ground-based high-resolution spectroscopy, Gaia astrometry, and cosmological zoom-in simulations can help to shed light on this question. Cosmological simulations suggest that VMP stars should be dispersed in a pressure-supported spheroidal distribution \citep[\aka the halo,][]{ElBadry18}, although the halo might naturally overlap with the disc in terms of space and kinematics, thus providing a similar contribution of prograde and retrograde stars.

VMPs with precise astrometric information were scarce before {\it Gaia} DR2 \citep{Gaia16,GaiaDR2}. It is now possible to complement the chemical information with precise astrometry, needed to derive distances and kinematical properties of the stars. Orbital properties are a useful tool for unveiling the multitude of accreted systems that concurred to form the Milky Way (MW) as we know it today \citep[\eg][]{Helmi20,Horta23}.  \citet{Sestito19} discovered that $26$ percent of known ultra-metal-poor stars (UMP, $\FeH\leq-4.0$) have kinematics confined to the MW plane (with a maximum height of $\lesssim3.5\kpc$), with 10 stars in prograde orbits and only one in retrograde orbits. Among the ones in prograde motion,  two are on nearly circular orbits \citep[see also][]{Schlaufman18,Mardini22,Dovgal24}, and the most metal-poor star known \citep[SDSS J102915+172927,][]{Caffau11} is one of these two. To answer whether this kinematical signature is also present at metallicities  $-4.0 \leq \FeH \leq -2.0$, \citet{Sestito20} gathered the VMPs in the solar neighbourhood from the LAMOST survey and from  Pristine targets with measured radial velocity \citep{Aguado19}, for a total of $\sim5000$ stars. They found that the majority of the stars are distributed in the halo; however, one third of VMPs are confined to the plane. 
Of these stars, prograde motion is favoured \citep{Sestito20}. Similar results have also been reported by the SkyMapper survey \citep{Cordoni21},  also showing that the [$\alpha$/Fe] ratios are compatible with the halo. Using the stars observed in the ESO Large Program "First stars program" \citep{Cayrel04,Bonifacio09}, \citet{DiMatteo20} confirmed that stars in planar orbits exist at all metallicities, suggesting that the dissipative collapse that led to the formation of the old and metal-poor Galactic disc must have been extremely fast. 

Both \citet{Sestito19} and \citet{Sestito20} propose that the prograde and retrograde planar  stars might have originated in the building blocks of the proto-MW or in a system accreted later during the MW history, while the prograde ones might also be the tracers of the MW disc's very metal-poor tail. The origin of the prograde planar population have been investigated also with the latest {\it Gaia} DR3 \citep{GaiaDR3}, since it is also releasing metallicities and radial velocities for millions of stars. Within the Pristine survey, \citet{Viswanathan25} found that the prograde region with low vertical angular momentum is overdense with a significance of $4\sigma$ more than its retrograde counterpart. \citet{ZhangArentsen24} found that the disc, as a major component, does not extend to the VMP regime, but, rather, that the VMP  prograde planar population is part of the rotating halo. In contrast, \citet{Bellazzini24} and \citet{Gonzalez24} propose that the angular momentum distribution of the prograde planar stars is similar to the more metal-rich disc and it might be of in-situ origin. \citet{Gonzalez24} discuss that this prograde planar population bears some of the properties that are classically associated with a thick disc, \ie its spatial distribution is compatible with a short scale-length thick disc and its presence even in case of assuming a rotating prograde halo. Moreover, \citet{FernandezAlvar21} found a population of VMP towards the Galactic anticentre that would be compatible with the MW thin disc,  and \citet{FernandezAlvar24} corroborate it in all directions. Recently, \citet{Nepal24} derived  stellar ages and distances for $\sim200,000$ stars showing that the metal-poor ones with orbits compatible with the thin disc are in fact old, \ie $>13$~Gyr. Finally, \citet{FernandezAlvar25} have identified very old ($12$~Gyr) metal-poor stars (total metallicity [Z/H]~$< -1$) in both thick and thin disc like orbits by inferring the age and metallicity distribution of stars in a very local volume from a new more sophisticated approach of a CMD-fitting analysis (the \textit{Chronogal} project).

With high-resolution cosmological zoom-in simulations,  NIHAO-UHD \citep{Buck20} and FIRE \citep{Hopkins18}, \citet{Sestito21} and \citet{Santistevan21}, respectively, investigated  the potential formation sites of this VMP planar population. Both  suites of simulated galaxies provide similar results. During the early Galactic assembly, the merging of the first building blocks scattered the stars in all  kinematical configurations, given the shallow gravitational potential of the forming proto-galaxies. Once the proto-Galaxy grew and  the Galactic disc started to form, later accreted systems deposited their stars mainly in the halo or, in some cases, into the plane  in a prograde fashion by dynamical friction \citep{Abadi03,Sestito21,Santistevan21}. In this picture, the retrograde planar stars can only originate from the early MW assembly phase. In these simulations, the galactic disc is detectable  after the formation of these stars, therefore, these VMP planar stars are not part of the VMP tail of the simulated disc. Recently, \citet{SotilloRamos23} used the TNG50 cosmological simulations \citep{Pillepich19,Nelson19} with a similar intent as the previous works. They found that the origin of the prograde planar population can also be attributed to in-situ formation. However, a complete census of detailed chemical abundances is still lacking in the literature. Having a better insight into the chemical properties of these stars would be beneficial to understand the merging history of the Galaxy, as well as the chemical evolution of the oldest part of the disc, as well as the chemical properties of the first generations of stars.

In this work, we investigate the chemical properties of VMP planar stars selected from \citet{Sestito20} and observed with the ESPaDOnS high-resolution spectrograph at the Canada-France-Hawaii Telescope (CFHT). We want to highlight that the stars in our sample have relatively large eccentricities, hence this work does not focus on stars with  quasi-circular orbits. The paper is organised as follows, Section~\ref{sec:data} describes the observations and their reduction; the calculations of the astrometric distances and orbital parameters are in Section~\ref{sec:orb}; the model atmosphere analysis is described in Section~\ref{sec:modelatm}; Section~\ref{sec:hidden} discusses the chemical properties of our targets, also in comparison with the Milky Way and dwarf galaxies; Section~\ref{sec:hints} discusses various tests that suggest the chemical peculiarity of this population;  Section~\ref{sec:chemevo} discusses whether a common origin for the planar stars is possible, also in  comparison with a larger  compilation of planar stars that have a wider range of kinematical properties. The summary of our findings is reported in Section~\ref{sec:conclusions}.

\section{Target selection and spectral reduction}\label{sec:data}

\begin{figure*}
\includegraphics[width=\textwidth]{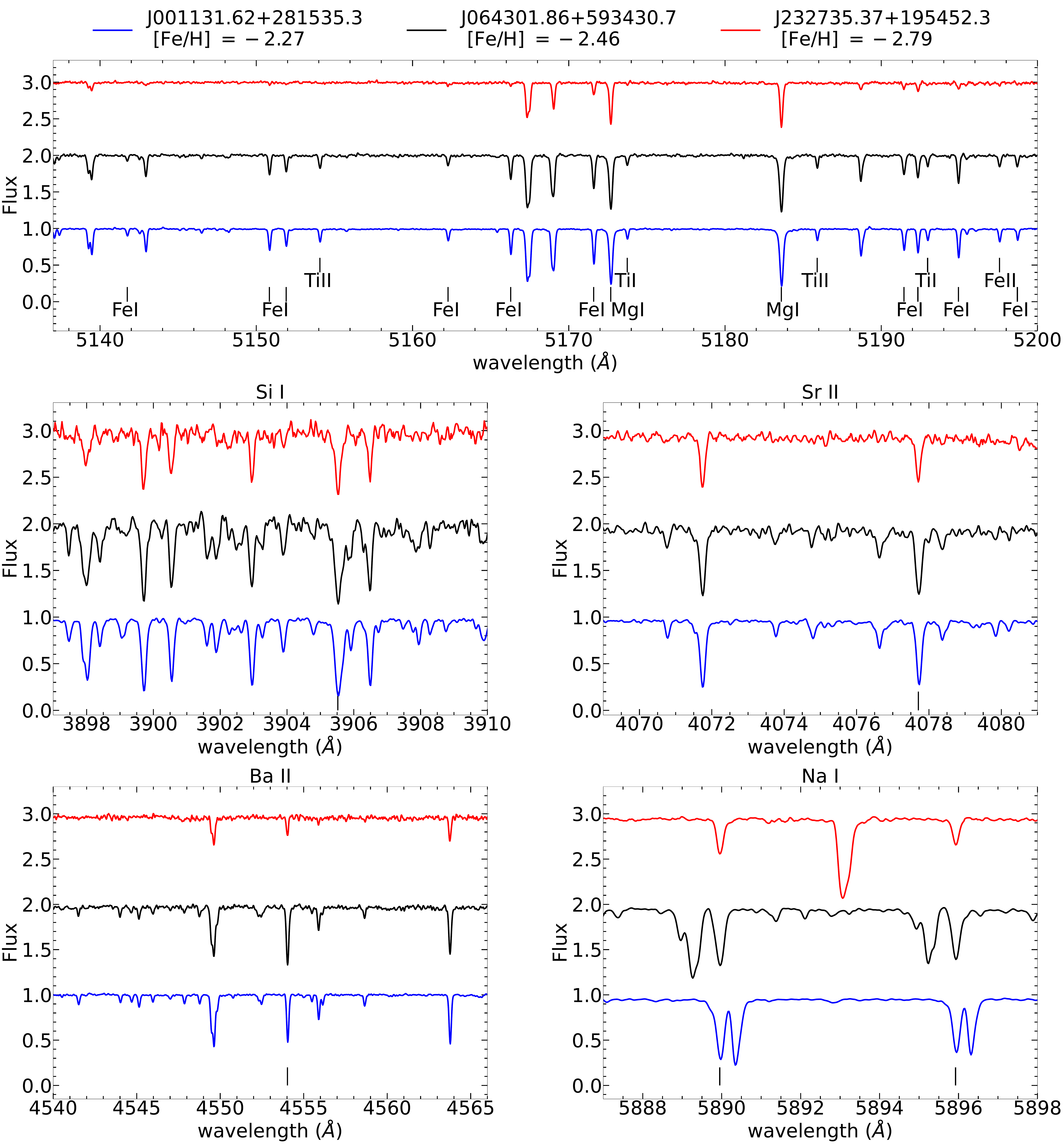}
\caption{Example of the ESPaDOnS spectra. Top panel: the Mg\ione{} Triplet region, which also includes several Fe and Ti lines. Vertical  short lines mark the position of Mg\ione{}, Ti, and Fe lines found in most of the targets. Central left panel: the Si\ione{} 3905.523 \AA{} line region. Central right: the Sr\ii{} 4077.709 \AA{} line region.  Bottom left: the Ba\ii{} 4554.029 \AA{} line region. Bottom right: the Na\ione{} Doublet $\lambda\lambda 5889.951, 5895.924$ \AA{} lines region, which also includes spectral lines from various clouds of the interstellar medium. The range of SNR measured in the Mg\ione{}b Triplet region of the whole of sample spans from 40 (a star observed in snapshot mode) to 220 (in normal mode).}
\label{Fig:spectra}
\end{figure*}

Targets used in this work were selected from \citet{Sestito20}, which combines very metal-poor stars (VMP, $\FeH\leq-2.0$) from the Pristine \citep{Starkenburg17b,Aguado19} and LAMOST \citep{Lamost12} surveys.  The initial sample of  stars has been selected using the action space of Figure~1 in \citet{Sestito20} (vertical vs rotational motion), and include stars that do not venture far from the Milky Way plane (Z$_{\rm{max}} \lesssim 3.5 \kpc$) in prograde and retrograde orbits. 
The second selection is based on the brightness, $10.0 \lesssim G \lesssim 13.0$ mag, to reach a high signal-to-noise ratio (SNR) in the blue region  within a reasonable exposure time at ESPaDOnS/CFHT. Given this  restriction on magnitudes, only stars from the LAMOST catalogue have been observed, given  stars from Pristine reported in \citet{Sestito20} are mostly fainter than G~$\sim14$ mag \citep{Aguado19}. Although the prograde planar stars are more numerous than the retrograde counterpart \citep{Sestito20}, the final selection is composed of an equal number of stars from the two populations.

The selected objects have been observed in semesters 2021B, 2022A, and 2022B (21BC23, 22AC37, and 22BC14, PI: F. Sestito) with the ESPaDOnS high-resolution spectrograph at the Canada-France-Hawaii Telescope (CFHT), which covers the $3700 - 10500$ \AA{} spectral region. Objects have been observed with the star$+$sky mode with a resolution of $\sim68000$.  2021B observations were carried out in Snapshot mode, leading to only 2 observed stars. Observations for the 2022A and 2022B semesters were carried out in regular mode, and the spectra of 9  stars were acquired during each semester. Therefore, the observed sample is composed of 20 unique stars. Spectra observed in Snapshot mode have a slightly lower SNR than the other data, given the lower quality of the sky conditions, which results in fewer species/lines measured in the bluest part of the spectrum. A SNR per CCD pixel bin of $\sim75$ in the Ba\ii{} 4554~\AA{} region was requested to ensure precision measurements of the equivalent width of elements such as Sr\ii{} (\eg $\lambda\lambda 4078, 4216$~\AA), Ba\ii{} (\eg $\lambda\lambda 4554, 4934, 5854, 6142$~\AA), Mn\ione{} (\eg $\lambda\lambda 4031, 4460, 4754, 4783$ \AA), and Eu\ii{} (\eg $\lambda\lambda 4130, 4205$~\AA). This SNR  permits a precision of $<0.15$ dex in the [X/Fe] ratios.

ESPaDOnS spectra are  reduced using the LIBRE-ESPIRIT\footnote{\url{https://www.cfht.hawaii.edu/Instruments/Spectroscopy/Espadons/Espadons_esprit.html}} pipeline, which includes bias subtraction, flat fielding, wavelength calibration, spectral extraction, and barycentric corrections. A second reduction and optimisation of the spectra was carried out using a dedicated pipeline  \citep{Lucchesi22}. The pipeline isolates and extracts the \'echelle orders, then it selects the regions with the highest SNR between the overlapping regions of the orders. The orders are then re-combined and re-normalised. Radial velocities, RV, have been measured with \textsc{fxcor} in \textsc{IRAF} \citep{Tody86,Tody93} and the spectra are corrected accordingly. Multiple exposures of the same target have been combined together and the combined spectrum has been downgraded to resolution $\sim40,000$  \citep[see][for further details]{Lucchesi22}. A table containing the coordinates, the {\it Gaia} info, the RV, and the measured SNR is reported as online material. Panels in Figure~\ref{Fig:spectra} show the high-quality spectra of three stars in various regions, namely the Mg\ione{} Triplet  ($5140-5200$~\AA, top panel),  the Si\ione{} $3905$~\AA{} (central left), the Sr\ii{} $4078$~\AA{} (central right), the Ba\ii{} $4554$~\AA{} (bottom left), and the Na\ione{} Doublet $5890,5896$~\AA{} (bottom right).

\section{Distances and orbital parameters}\label{sec:orb}

Distances are calculated using the exquisite {\it Gaia} DR3 parallaxes in a Bayesian fashion. The posterior, or the probability distribution function, is obtained by multiplying the Gaussian likelihood on the parallax times a Galactic  stellar density distribution prior \citep[see][for further details]{Sestito19}. The zero point offset has been applied to the {\it Gaia} DR3 parallaxes \citep{Lindegren21} using the python \textsc{gaiadr3\_zeropoint}\footnote{\url{https://gitlab.com/icc-ub/public/gaiadr3\_zeropoint}} package.

\begin{figure*}
\includegraphics[width=0.3\textwidth]{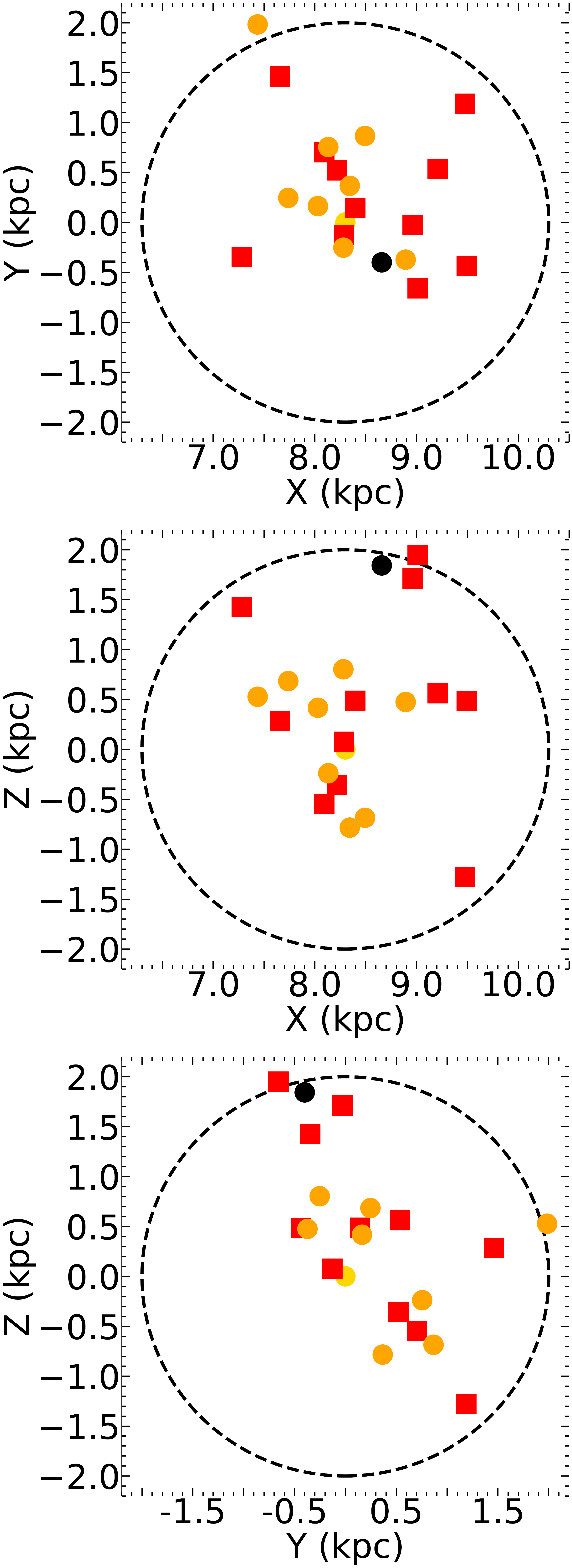}
\includegraphics[width=0.67\textwidth]{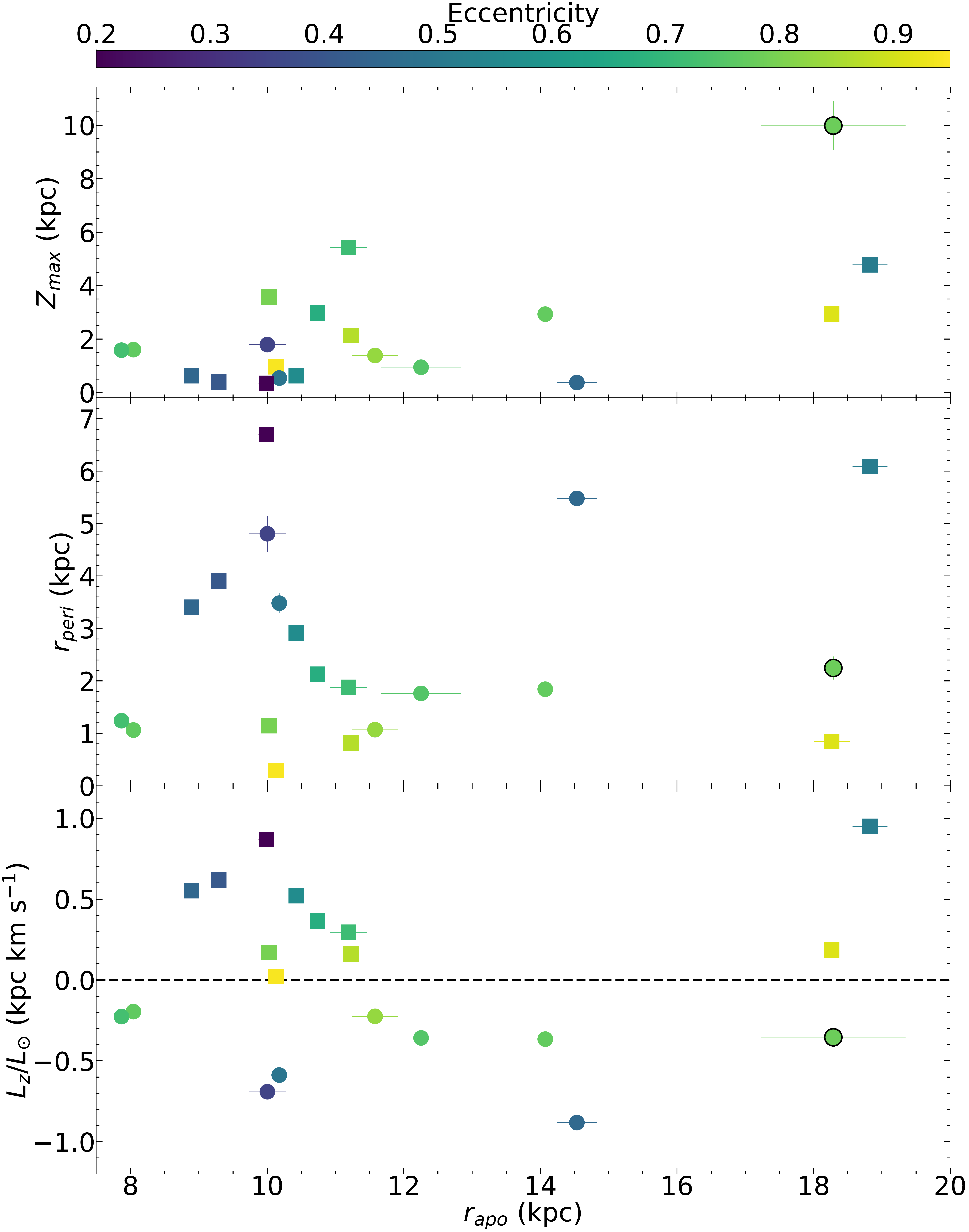}
\caption{Orbital  parameters. Left panels: Current Galactocentric positions, Y vs X (top left), Z vs X (centre left), and Z vs Y (bottom left). Red squares and orange circles denote the prograde and retrograde targets, respectively. J110847.18$+$253047.2 is marked with a black circle (the star with the highest maximum height from the plane). Green circle represents the position of the Sun, (X, Y, Z)$_{\odot}$~$=$~($8.3, 0.0 , 0.0$) kpc. Right panels: Maximum height from the plane (top right), pericentric distance (centre right), and z-component of the angular momentum (bottom right) vs the apocentric distance. The angular momentum is normalised by the Sun's value, L$_z = 2010 \kpc \kms$. Markers are colour-coded by the eccentricity. Squares and circles indicate prograde and retrograde orbits, respectively. J110847.18$+$253047.2's marker edge is a black circle.}
\label{Fig:kine}
\end{figure*}

The kinematical information has been updated from {\it Gaia} DR2 \citep[values from][]{Sestito20} to {\it Gaia} DR3 values, including new astrometric distance and  new radial velocities measured from  ESPaDonS spectra.
Orbital parameters are inferred using \textsc{galpy} \citep{Bovy15} and adopting the same Galactic gravitational potential as in \citet{Sestito19}. The potential is a modified version of \textsc{MWPotential2014} with an increased dark matter halo mass of $1.2 \times 10^{12}\rm M_{\odot}$ \citep[][]{BlandHawthorn16}. Monte Carlo simulations have been used to generate uncertainties on the orbital parameters from the input parameters (distance, RV, proper motion, coordinates), drawing them from  Gaussian distributions for 1000 times. The median and the standard deviation are used as the measurement of an orbital  quantity and its uncertainty. 

The current Galactocentric positions of these stars are displayed in the left panels of Figure~\ref{Fig:kine}.  The maximum height from the plane $\rm{Z_{max}}$, the angular momentum $\rm{L_{z}}$ (normalised by the Sun's value) and the Galactic pericentric distance $\rm{r_{peri}}$ are displayed in the right panels of Figure~\ref{Fig:kine} as a function of the Galactic apocentric distance $\rm{r_{apo}}$. 
While the current position of these stars is within $\lesssim2.3$ kpc from the Sun, 11 of them reach the inner Galactic region ($\lesssim3$ kpc from the Galactic centre) at their pericentric passage, while the apocentric distance spans a range from $\sim8$ to $\sim18$ kpc. The eccentricities of these stars span a range from $\sim0.20$ to $0.95$. The majority of the stars reaches a maximum height from the plane $\rm{Z_{max}}\lesssim3.5$ kpc, with one kinematical outlier in $\rm{Z_{max}}$ at $\sim10$ kpc, J110847.18$+$253047.2 (black circle in Figure~\ref{Fig:kine}). This object is also the only star with a large uncertainty on the radial velocity $\sim9\kms$ (vs $\sim2\kms$). The numbers of stars with prograde and retrograde motion are 11 and 9, respectively.
While these stars were originally selected to have $\rm{Z_{max}}<3.5\kpc$, the updated astrometry from {\it Gaia} DR3 and the new RV from high-resolution spectra lead to  higher $\rm{Z_{max}}$ for some stars. However, we will discuss in Section~\ref{sec:hidden} that  J110847.18$+$253047.2 is the only chemical outlier. The kinematical properties and heliocentric distances are reported as an online table only.

\section{Model Atmosphere analysis}\label{sec:modelatm}

\subsection{Stellar parameters}\label{sec:stellparams}

\begin{figure}
\includegraphics[width=0.5\textwidth]{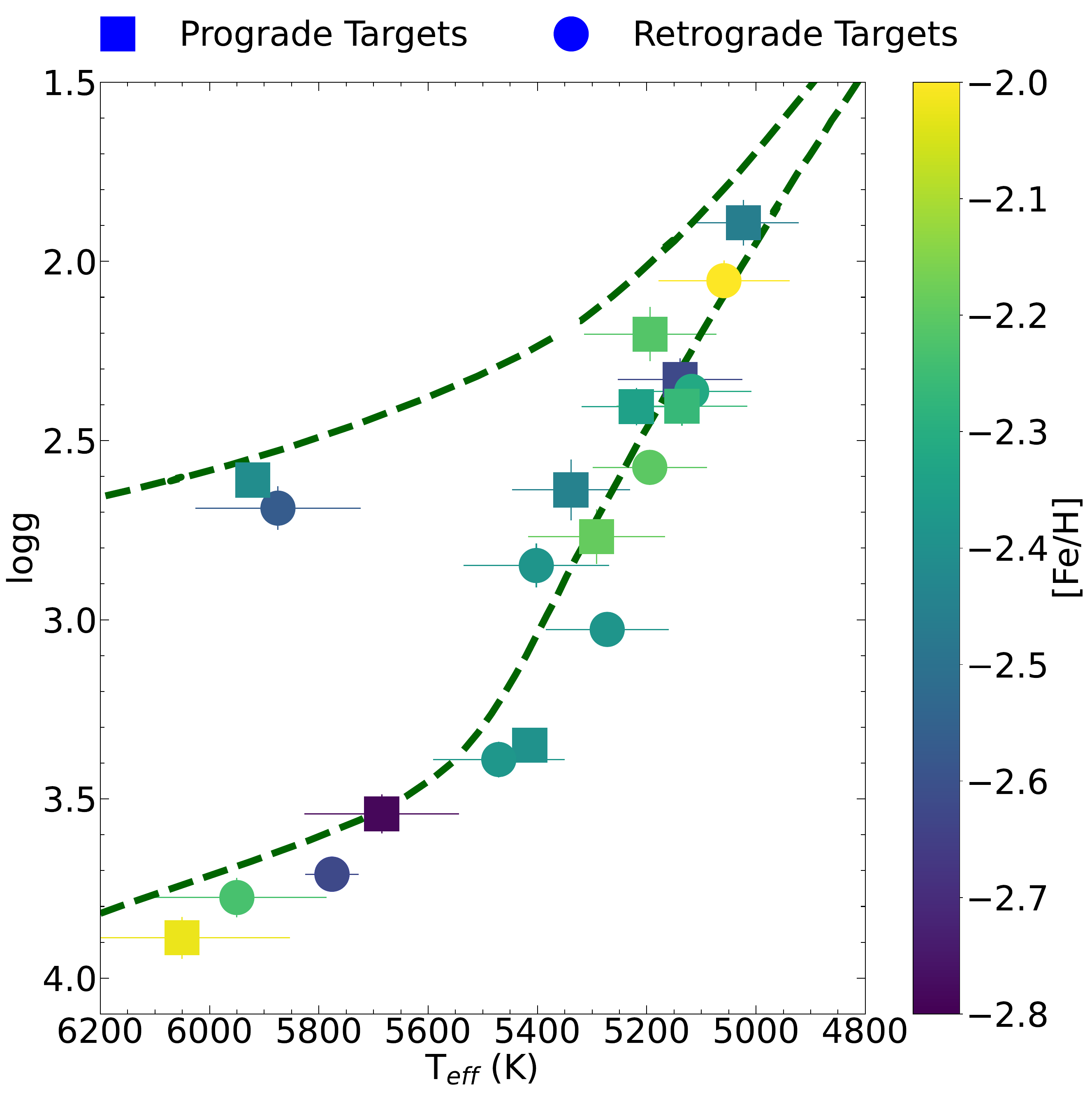}
\caption{Kiel diagram. Targets are colour-coded by metallicity. Retrograde and prograde stars are denoted by squares and circles, respectively. A Padova isochrone \citep{Bressan12} of $\FeH = -2.0$ (dark olive dashed line) is shown for comparison.}
\label{Fig:stellar}
\end{figure}

The effective temperature, $\teff$,  is derived from  a colour-temperature relation similar to the Infrared Flux Method \citep{Gonzalez09} and adapted to the {\it Gaia} DR3 photometry \citep{Mucciarelli21}. The surface gravity, $\log g$, is determined by applying the Stefan-Boltzmann equation \citep[\eg][]{Kraft03,Venn17}. An iteration between $\teff$ and $\log g$ is used, following the method fully described in \citet{Sestito23}. Uncertainties on the stellar parameters are derived with a Monte Carlo simulation on  the input parameters \citep[photometry, distance, and extinction from][]{Schlafly11}.  The Kiel diagram is displayed in Figure~\ref{Fig:stellar} with a very metal-poor Padova isochrone \citep{Bressan12} for comparison. The plot shows that the sample is composed of 4 sub-giants, 14 red giants, and 2 horizontal branch (HB) stars. The microturbulence velocity (v$_{\rm{micro}}$) is derived spectroscopically, imposing a flat relation between abundances of Fe\ione{} and the reduced equivalent width. The stellar parameters and their uncertainties are reported as online material only.

\subsection{Spectral line list, atomic data, and model atmosphere}
The line list and the atomic data are sourced from \textsc{linemake} \citep{Placco21}, which includes hyper-fine structure (HFS) and r-process isotopic (for Ba and Eu) corrections.   
Molecular CH bands are from \citet{Masseron14}. Solar abundances are adopted from \citet{Asplund09}.  The high SNR  of these spectra and the wavelength coverage of ESPaDOnS allow us to detect atomic lines of $\alpha$- (Mg, Si, Ca, and Ti), odd-Z (Na, Al, K, Sc, and V), Fe-peak (Cr, Mn, Co, Ni, and Zn), and neutron-capture (Sr, Y, Zr, Ba, La, Nd, and Eu) elements and molecular bands of CH. The  line list and their atomic are reported as supplementary online material.

\begin{figure}
\includegraphics[width=0.5\textwidth]{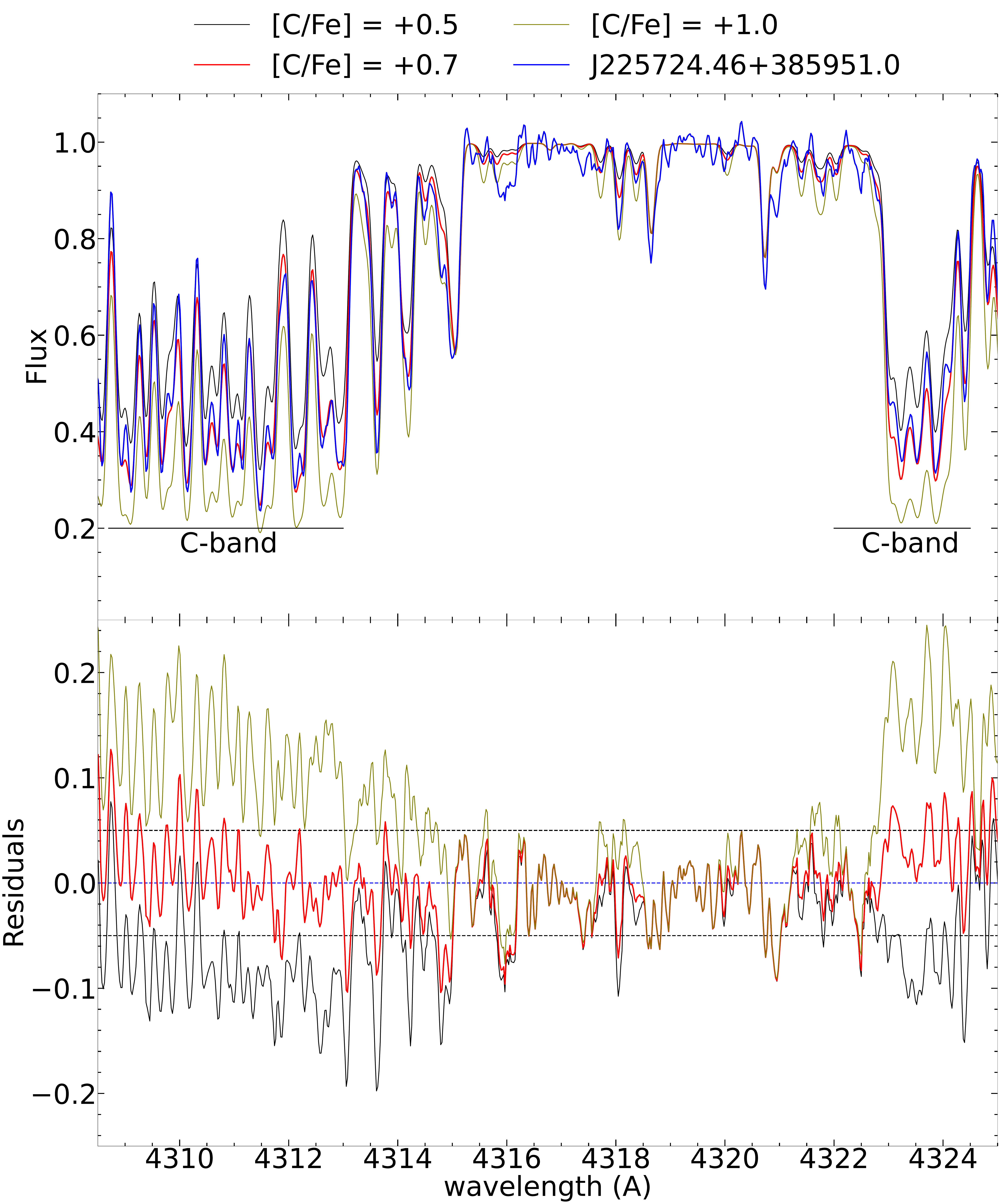}
\caption{Carbon synthesis. Top panel: The ESPaDOnS spectrum of J225724.46$+$385951.0 is represented by the blue line, while synthetic spectra are shown by the black ([C/Fe] $=+0.50$), red ([C/Fe] $=+0.7$, best fit), and olive ([C/Fe] $=+1.00$) lines. Bottom panel: Residuals of the fits. The horizontal dotted lines mark the null difference (blue) and the dispersion around the continuum ($\pm 0.05$ dex, black). The residuals of the best fit (red line) in the regions of the C bands are within or similar to the level of the continuum dispersion.}
\label{Fig:carbon}
\end{figure}

The equivalent width (EW) of the spectral lines is measured using \textsc{DAOSPEC} \citep{Stetson08}, which automatically fits the lines with a Gaussian profile. The EWs of very strong lines have been revised using a Voigt profile. Given the signal-to-noise ratio of our spectrum,  lines weaker than 15 m\AA{} are rejected.  Chemical abundances are derived with the \textsc{MOOG}\footnote{\url{https://www.as.utexas.edu/~chris/moog.html}} code \citep{Sneden73,Sobeck11}  adopting the \textsc{MARCS}\footnote{\url{https://marcs.astro.uu.se}} model atmospheres \citep{Gustafsson08,Plez12}. EWs and derived elemental abundances are reported as supplementary online material. Carbon abundance is measured using the \textsc{synth} mode in \textsc{MOOG}, from the CH $\sim4300$ \AA{} band \citep{Masseron14}, assuming an isotopic ratio of $\rm{^{12}C/^{13}C} = 20$ \citep{Spite06}. Evolutionary corrections to [C/Fe] have been applied following \citet{Placco14}\footnote{\url{https://vplacco.pythonanywhere.com}}. As an example for C-synthesis, the top panel of Figure~\ref{Fig:carbon} shows the spectrum of J225724.46$+$385951.0 and three synthetic spectra ([C/Fe]~$+0.5,+0.7,+1.0$), while the  residuals are displayed in the bottom panel. The star is further discussed in Section~\ref{sec:tree_out} as it is enhanced in C ([C/Fe]~$=+0.7$) and Ba ([Ba/Fe]~$\sim+1.4$). 

The atmosphere of VMP stars are affected by non-local thermodynamic equilibrium (NLTE) effects, which can be large for some species. 
NLTE corrections have been applied to Fe\ione{} and Fe\ii{} \citep{Bergemann2012}, Mg\ione{} \citep{Bergemann2017}, Si\ione{} \citep{Bergemann2013}, Ca\ione{} \citep{Mashonkina17}, Ti\ione{} and Ti\ii{} \citep{Bergemann2011}, Cr\ione{} \citep{Bergemann2010b}, Mn\ione{} \citep{Bergemann2019}, and Co\ione{} \citep{Bergemann2010} using the MPIA data base\footnote{\url{https://nlte.mpia.de}}. For Na\ione{} \citep{Lind2012} and Sr\ii{} \citep{Bergemann2012b},  the \textsc{INSPECT}\footnote{\url{http://www.inspect-stars.com}} webtool has been used.  Ba\ii{} NLTE corrections are adopted from \citet{Mashonkina19} taken from their online database\footnote{\url{http://www.inasan.ru/~lima/pristine/ba2/}}.  NLTE corrections for Al and K are obtained from \citet{Nordlander17b} and \citet{Ivanova00}, respectively.  A table containing NLTE-corrected chemical abundances is provided as supplementary online material.

\subsection{Uncertainties on the chemical abundances}
\textsc{MOOG} provides estimates of the chemical abundances A(X) along with their dispersion, $\delta_{\rm A(X)}$. The values of the chemical abundance dispersions have been reduced by the square root of the number of lines of a given species, $\delta_{\rm A(X)}/\sqrt{{\rm N_X}}$. Then, these values have been summed in quadrature with the uncertainties resulting from variations in the stellar parameters ($\delta_{\rm T_{eff}}$, $\delta_{\rm logg}$, $\delta_{\rm{v_{micro}}}$) to obtain the final uncertainty for a given element, $\sigma_{\rm A(X)}$. When only one spectral line is present, the dispersion is set to be similar to the dispersion of the Fe\ione{} lines in the same spectral region.

\section{Chemical properties of the sample}\label{sec:hidden}

\subsection{Chemical abundances with respect to the Milky Way halo}\label{sec:comparisonhalo}

\begin{figure*}
\includegraphics[width=\textwidth]{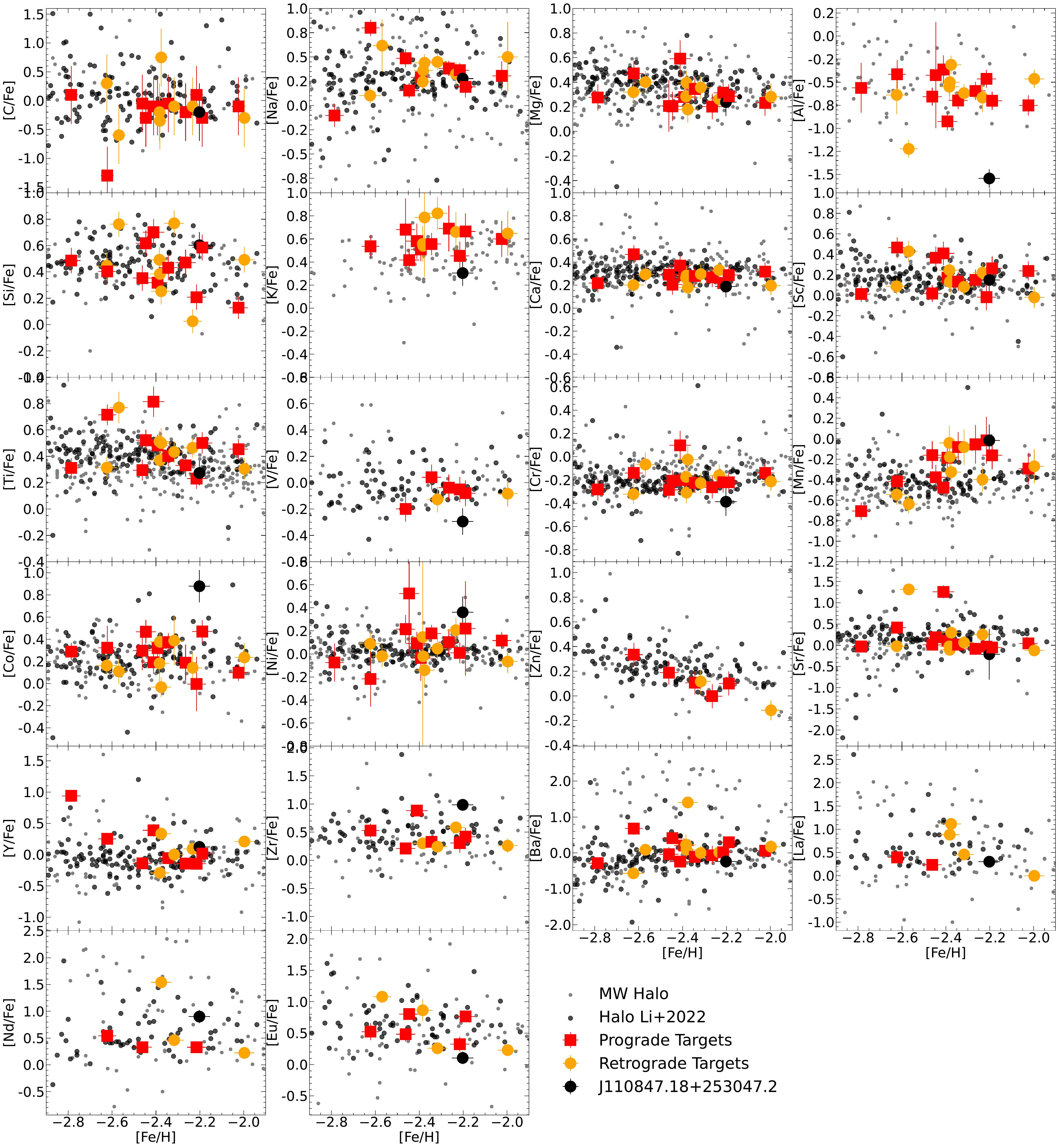}
\caption{Chemical abundances as a function of [Fe/H]. Prograde and retrograde targets are marked with red  squares and orange circles, respectively. MW halo from the SAGA database \citep{Suda08} and \citet{LiH22} are denoted by  smaller and larger grey circles, respectively. All datasets are in LTE. Halo stars have been selected to have a similar distribution  in the stellar parameters as our targets for a more robust comparison.}
\label{Fig:chems}
\end{figure*}

The quality of the observed ESPaDOnS spectra allow us to measure the chemical abundances of many light and heavy elements. The chemical abundances of our sample are compared to a compilation of stars in the Milky Way. This compilation is composed of high-resolution observations extracted from the Stellar Abundances for Galactic Archaeology database\footnote{\url{http://sagadatabase.jp}} \citep[SAGA,][]{Suda08} and from the VMP sample analysed in  \citet{LiH22}. 
The latter sample is homogeneously observed with SUBARU/HDS and has a  similar spectral resolution ($\sim36,000$) as our observations. 
Literature stars with similar stellar parameters as our targets ($\log g <3.8$, $\rm{T_{eff} <6000}$K) are selected to minimise  discrepancies in the NLTE corrections between giants and dwarfs. The comparison of the LTE chemical abundances is reported in Figure~\ref{Fig:chems}, where the panels are sorted in order of increasing proton number. 

The general distribution of the [X/Fe] ratios in our sample is in good agreement with the MW halo, embedded in the global distribution, although a simple visual inspection already reveals that our sample has a narrower chemical spread than the rest of the halo distribution for several elements, and that there are chemical outliers that might not be representative of the general population.  The C-panel of Figure~\ref{Fig:chems} shows that our sample is located at the low end of the MW halo stars' distribution, with the exception of J225724.46$+$385951.0. This object is the only C-enhanced star in the observed sample.
The percentage of C-enhanced metal-poor stars (CEMP) in the VMP MW halo is still an open question and it is strongly biased by the different selection effects of spectroscopic surveys \citep{Arentsen22}. However, the cumulative percentage is around  $\sim20-30$ percent for stars with $\FeH<-2.0$ \citep{Placco14,Arentsen22}, which is  higher than the one in our sample ($\sim5$ percent). Given the low number of stars in our sample and our restrictive selection criteria, this result has to be taken with the grain of salt. While [C/Fe] is not a parameter of our target selection, it is unknown if the LAMOST survey is biased against CEMP stars. CEMP can be used as an excellent tracers for binarity (CEMP-s) and for supernovae signatures (CEMP-no) \citep[\eg][]{Beers05,Aoki07}. Recently, \citet{Lucchesi24} discussed that the percentage of CEMP-no stars in classical dwarf galaxies (DGs) is lower than the MW halo's and the ultra-faint dwarfs' (UFDs) fraction. Similarly, \citet{Sestito24Sgr2} found that Sagittarius, and various DGs, has a lower average [C/Fe] compared to the MW. \citet{Sestito24Sgr2} interpreted these results as the chemical imprint of a different population of supernovae between DGs and the MW halo. Classical DGs might have been able to retain the yields from the most energetic events, which produced more Fe than C, while the smaller building blocks of the MW halo and ultra-faint dwarfs would have been able to recycle material from faint events only \citep[\eg][]{Ji16c,Roederer16b,Hansen17,Kobayashi20,Applebaum21,Koutsouridou23,Vanni23,Waller23,Sestito24gh,Skuladottir24}. Similarly, the formation site(s) of these targets might contain the imprint of high energy supernovae events, similarly to a DG. 

Figure~\ref{Fig:chems} shows that the $\alpha$-elements are in good agreement with the MW halo distribution, and they do not display the sign of contribution from Type Ia supernovae, given the lack of an $\alpha$-knee (\ie a change in the slope of [$\alpha$-element/Fe] as a function of [Fe/H]). Regarding the odd-Z elements, they are in good agreement with the MW halo, however, [K/Fe] is higher in our sample. We note that J110847.18$+$253047.2 (large black circle) is strongly Al-poor and also lower in K and V than the other targets. This star also has a halo-like kinematics, rather than orbiting close to the plane as the other targets.  
[Mn/Fe] increases with metallicity and this ratio is slightly enhanced compared to the bulk of the MW values for stars with $\FeH\gtrsim-2.4$. The same negative slope is seen for [V, Cr, Zn/Fe] vs \FeH{} in both our sample and the MW halo. 
The panels of Figure~\ref{Fig:chems} relative to the neutron-capture processes (from Sr to Eu), reveal that the [X/Fe] of our targets are similar to those in the MW halo, with the exception of two Sr-rich stars and one Ba-rich star.

\subsection{Chemical similarities with classical dwarf galaxies}\label{sec:dgufd}

\begin{figure*}
\includegraphics[width=\textwidth]{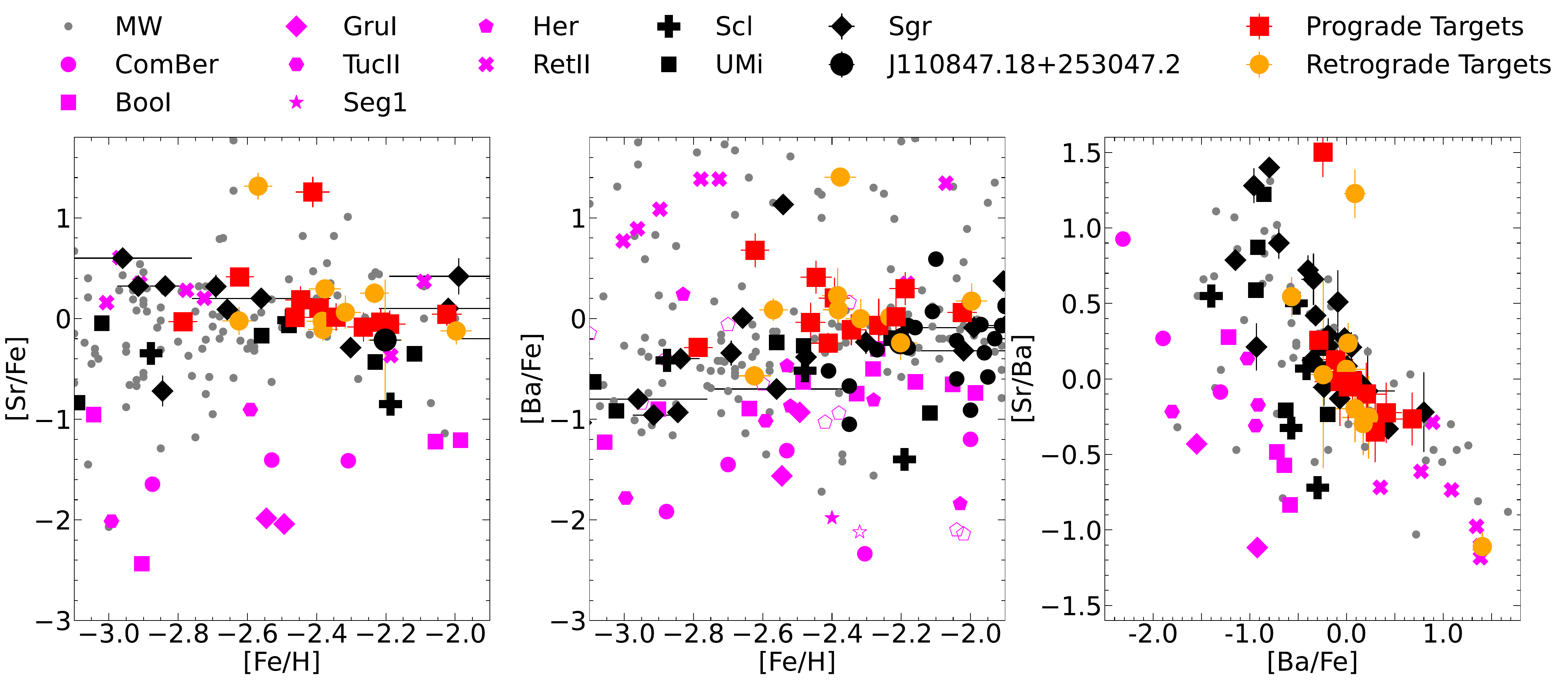}
\caption{Neutron-capture elements. Left panel: [Sr/Fe] vs \FeH{}. Central panel: [Ba/Fe] vs \FeH{}.  Right panel: [Sr/Ba] vs [Ba/Fe]. Empty symbols denote upper limits on the vertical axis. MW stars (grey circles) are from  SAGA  \citep{Suda08}; Coma Berenice (ComBer) stars are from \citet{Frebel10b} and \citet{Waller23}; Bootes~I (BooI) stars are from \citet{Feltzing09}, \citet{Norris10}, \citet{Gilmore13}, \citet{Ishigaki14}, and \citet{Frebel16}; Gru~I  stars are from \citet{Ji19}; Tucana~II (TucII) stars are from \citet{Ji16} and \citet{Chiti18}; Segue~1 (Seg1)  stars are from \citet{Frebel14}; Hercules (Her) stars are from \citet{Koch08}, \citet{Koch13}, and \citet{Francois16}; Reticulum~II (RetII) are from \citet{Ji16b} and \citet{Roederer16}; Sculptor (Scl) are from \citet{Mashonkina17b} and \citet{Hill19}; Ursa Minor (UMi) stars are from \citet{Mashonkina17b} and \citet{Sestito23Umi}; Sagittarius (Sgr) stars are from \citet{Hansen18Sgr} and \citet{Sestito24Sgr}. Magenta and black markers denote UFDs and DGs, respectively. Data are from 1D LTE optical analyses. The Sr\ii{} lines detected in our targets are  $\lambda\lambda 4078, 4216$~\AA{}, while those for  Ba\ii{} are $\lambda\lambda 4554, 4934, 5854, 6142$~\AA.} 
\label{Fig:neutron}
\end{figure*}

Neutron-capture elements and their ratios have been used to separate massive and chemically evolved DGs from more pristine  UFDs \citep[\eg][]{Mashonkina17b,Ji19}.  The content of Sr and Ba is on average lower in UFDs than in classical DGs. The distributions of [Sr, Ba/Fe] vs \FeH{}  are displayed in the two leftmost panels of Figure~\ref{Fig:neutron} for our targets (red and orange markers) and a compilation of stars from DGs (black markers), UFDs (magenta markers), and MW halo stars (grey circles). The Ba and Sr contents of our sample clearly appears to be similar to the average content of DGs, rather than of UFDs.

The right panel of Figure~\ref{Fig:neutron} shows [Sr/Ba] vs [Ba/Fe]. In this space, MW halo stars display a downward trend \citep{Mashonkina17b}, although distributed mostly around [Sr/Ba]~$\sim0.3$ \citep{Ji19}. Stars of a given UFD or DG exhibit a relatively wide distribution in [Sr/Ba], up to $\sim1.5$ dex. This can be explained by stochasticity in the interstellar medium (ISM) of UFDs and multiple nucleosynthetic channels in both UFD and DGs. Given the lower amount of Ba and Sr, UFDs populate a distinct region of this space for the majority of MW halo stars and of DGs. 

A ratio [Sr/Ba]~$\sim0$ implies that these species were produced by the same nucleosynthetic sources \citep{Mashonkina17b}, which seems to be the case for most of our targets. The three exceptions are the CEMP-s target J225724.46$+$385951.0, where Ba is likely enhanced by an AGB donor ([Ba/Fe]~$\sim+1.5$); and the two HB stars, J080626.72$+$194212.2 and J103037.10$+$224124.4. A high [Sr/Ba] ratio can be explained by an extra nucleosynthetic channel for Sr, which is still an open question and can be related to s-processes or weak r-process production, which would not involve synthesis of Ba \citep[see][and references therein]{Mashonkina17b,Sitnova25}.
In the [Sr/Ba] vs [Ba/Fe] space, our sample is clearly detached from  UFDs, resembling more the properties of stars in DGs and in the halo.

\subsection{Rapid and slow process sources}\label{sec:rsproc}

\begin{figure}
\includegraphics[width=0.47\textwidth]{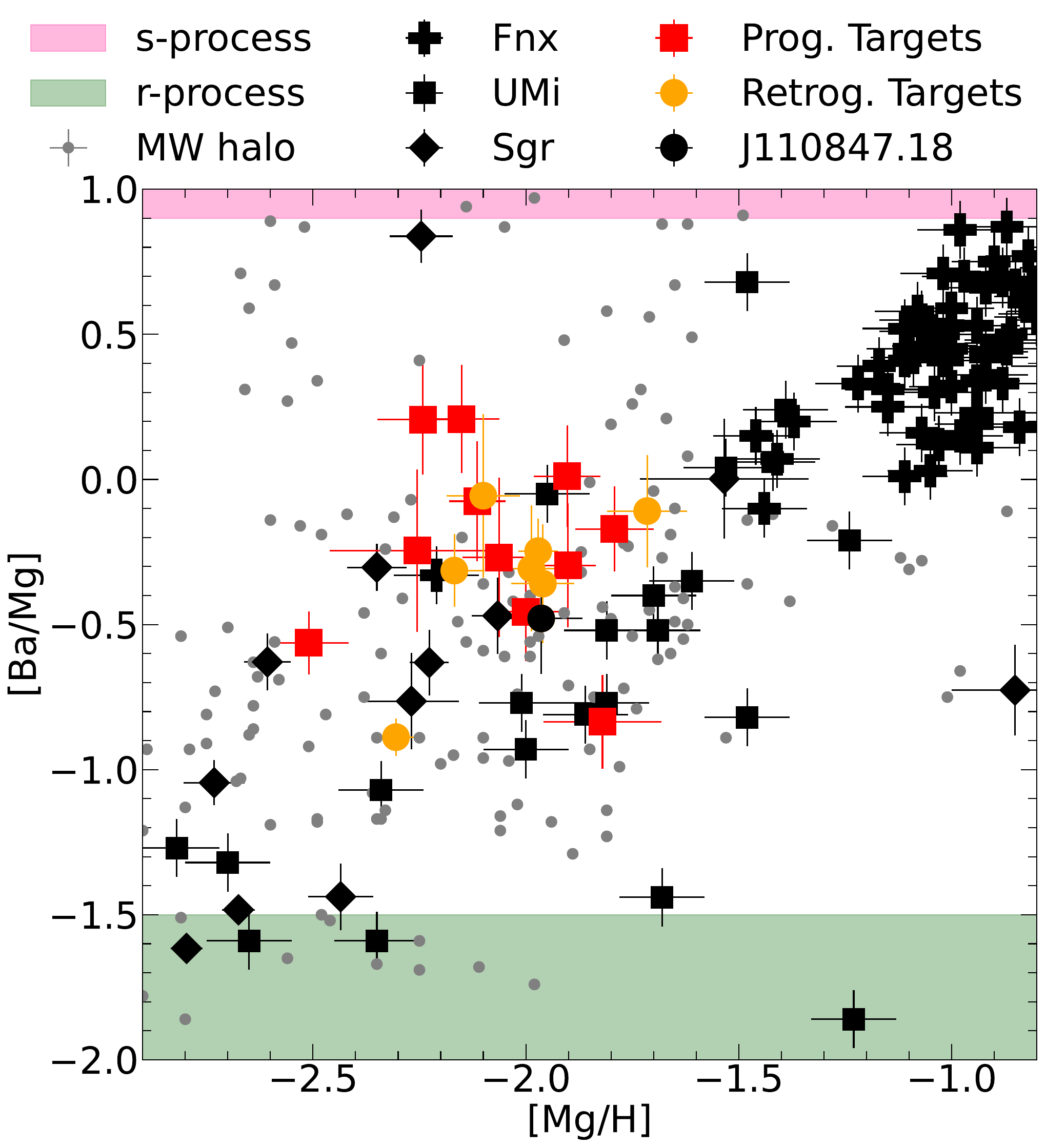}
\includegraphics[width=0.47\textwidth]{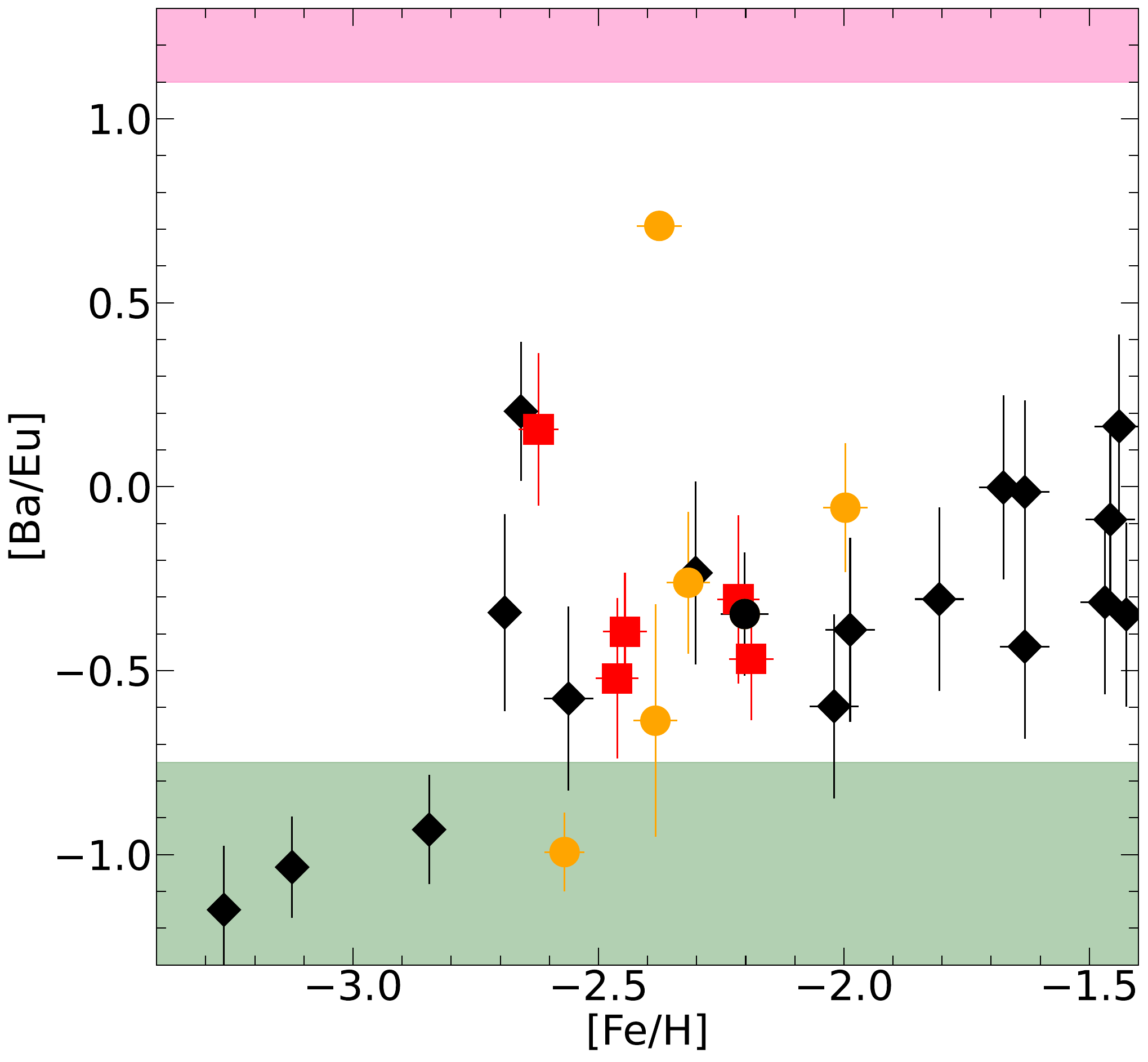}
\caption{Rapid and slow processes. Top panel: [Ba/Mg] vs [Mg/H].  Fnx stars (black squares) are from \citet{Letarte10}; UMi stars (black circles) are from \citet{Shetrone01}, \citet{Sadakane04}, \citet{Cohen10}, \citet{Kirby12}, \citet{Ural15}, and \citet{Sestito23Umi}; Sgr stars are from \citet{Hansen18Sgr} and \citet{Sestito24Sgr}. MW halo stars (grey circles) are from SAGA. Bottom panel: [Ba/Eu] vs [Fe/H]. The green area delimits the region where [Ba/Mg] is purely produced by r-process sources \citep[SNe~II yields from][]{Ebinger20}, while the pink shaded area indicates the region where [Ba/Mg] is produced by s-process sources \citep[AGB yields from][]{Straniero06,Cristallo07,Cristallo09,Arlandini99,Bisterzo14}. Lines of Ba\ii{} used in these chemical spaces are those at $\lambda\lambda 4554, 4934, 5854, 6142$~\AA, while those for Eu\ii{} are $\lambda\lambda 4130, 4205$~\AA. Elemental abundances of Mg\ione{} are measured from up to 10 lines from 3900 to 5720 \AA{}.}
\label{Fig:bamg}
\end{figure}

Eu and Mg are mostly synthesised by neutron stars mergers (NSM) and SNe~II, respectively, which are the main sources for rapid-capture process elements. Ba can also be made in slow process sources, namely fast-rotating and AGB stars. Therefore, [Ba/Mg] vs [Mg/H] and  [Ba/Eu] vs [Fe/H] chemical spaces are useful diagnostics to investigate contribution of both rapid and slow process sources. These chemical spaces are shown in Figure~\ref{Fig:bamg}, in the top and bottom panel, respectively.

The [Ba/Mg] vs [Mg/H] is reported for our sample  (orange circles and red squares), MW halo stars (grey markers), and  classical DGs (black markers), namely Fornax (Fnx), Ursa Minor (UMi) and Sagittarius (Sgr). The [Ba/Eu] vs [Fe/H] is displayed for our targets and Sgr members. In both panels, literature theoretical yields from pure rapid (green shaded area) and from slow processes (pink shaded area) are reported (see caption for references). Both [Ba/Mg] and [Ba/Eu] rise as a function of  [Mg/H] and \FeH{}, implying the need of an extra (s-process) source for Ba.

 \textsc{StarFit}\footnote{\url{https://starfit.org}}  is employed to fit the chemical pattern of our stars. \textsc{StarFit} takes as input the [X/Fe] with their uncertainties and provides the best combination of SNe~II through a $\chi^2$ fit. Theoretical supernovae and massive stars yields are selected to account for a variety of sources, from faint to pair instability SNe, to fast-rotating massive stars.

\textsc{StarFit} suggests that fast-rotating massive stars and compact binary merger events are needed to reproduce the observed neutron-capture elements in our sample. 
Additionally, high-energy supernovae and hypernovae are required to account for the production of the yields from  the $\alpha-$elements (high-energy and hypernovae) to the Fe-peak group (hypernovae only).
The imprint of the most energetic events vs the lack of faint- and core-collapse SNe has been recently proposed as typical of classical DGs \citep[\eg][]{Skuladottir24b,Sestito24Sgr2}. These systems should be able to retain the most energetic events and homogeneously recycle their yields, while the gas of UFDs would  be able to retain the ejecta of only  fainter events \citep[\eg][]{Ji16c,Roederer16b,Hansen17,Kobayashi20,Applebaum21,Waller23,Skuladottir24b,Sestito24gh,Sestito24Sgr2}.

\subsection{In-situ vs accreted origin}\label{sec:inexsitu}

The early Galactic assembly is a chaotic phase composed of the merging of the building blocks into the growing proto-Galaxy and, likely, an in-situ stellar formation. Traces of the latter have been found among stars with metallicities $-2.0<\FeH<-1.3$, based on their distinct chemical composition \citep{Belokurov22}. Therefore, the next question to address is whether the chemical properties of our sample provide evidence for an accreted origin or indicate that it formed in situ within the Galaxy. We want to emphasise that our definition of "accreted" includes stars either formed 1) outside the main host halo during the early Galactic assembly, \ie in the building blocks, or 2) in systems accreted into the MW at later time.

\begin{figure}
\includegraphics[width=0.5\textwidth]{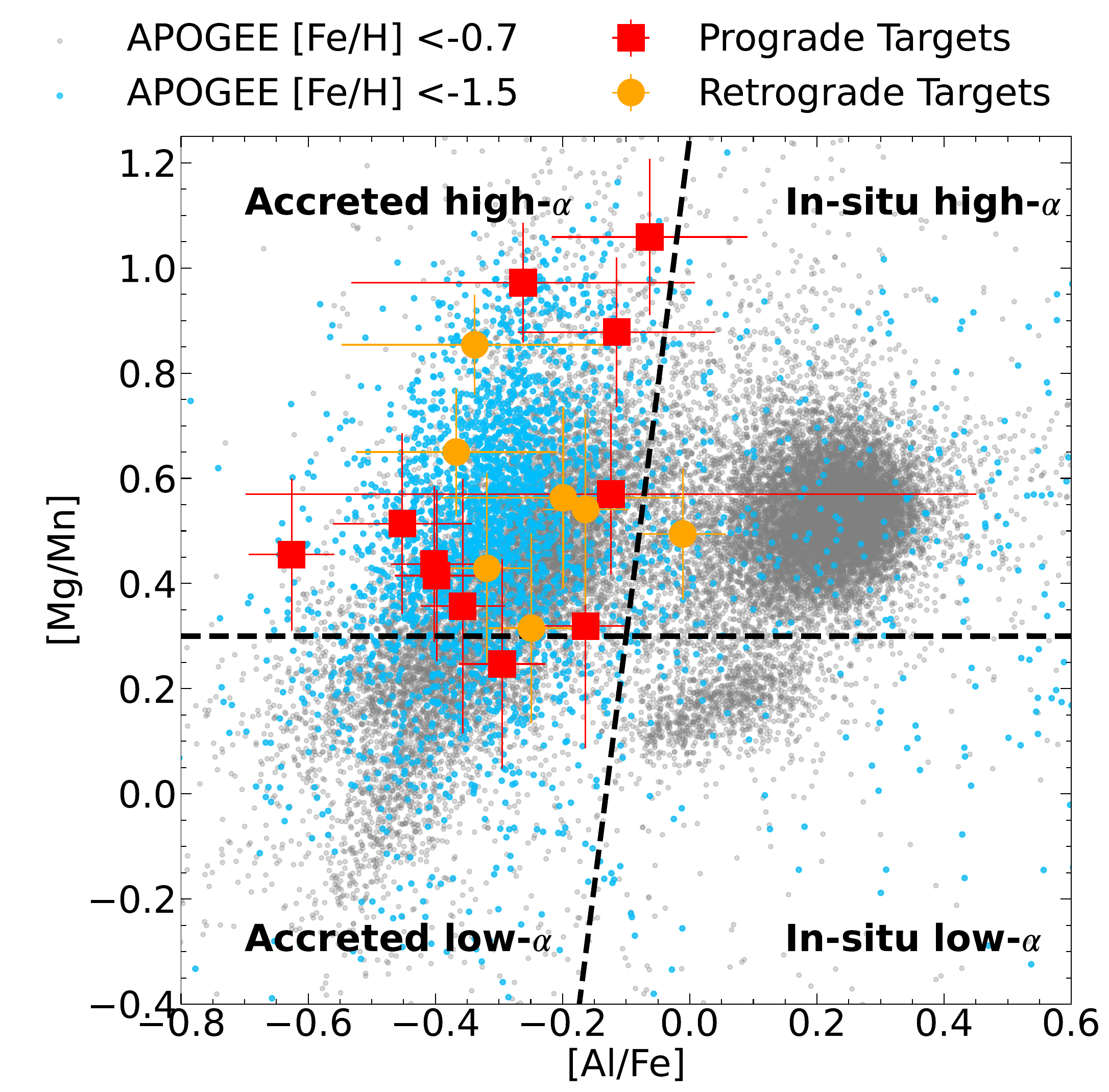}
\caption{[Mg/Mn] vs [Al/Fe] space. Our targets are marked as red squares (prograde) and orange circles (retrograde). J110847.18$+$253047.2 is not shown as it has a [Al/Fe]~$\sim-1.5$. Grey dots are APOGEE DR17 MW stars selected to have SNR $>70$ and $\FeH<-0.7$  (NLTE infra-red analysis). Light blue dots are APOGEE DR17 stars with the same quality cuts as the grey dots, but restricting to stars with $\FeH<-1.5$. The chemical abundances of our sample have been corrected for NLTE effects. Offsets between the NLTE infra-red (APOGEE) and the NLTE-corrected optical analyses are estimated to be $\Delta\rm{[Mg/Mn]}= +0.15\pm 0.18$ and $\Delta\rm{[Al/Fe]}= -0.11 \pm 0.12$ \citep{Jonsson20}, which are applied to our sample. The 4 regions delimited by the dashed black lines, accreted/in-situ low-/high-$\alpha$ are defined following \citet{Hawkins15}, \citet{Das20}, and \citet{Sestito24gh}. Chemical abundances for Al are measured from the Al\ione{} doublet $\lambda\lambda3944, 3961$~\AA{}, while those for Mn  are usually from the Mn\ione{} Triplet at $\sim4033$~\AA{} and from $\lambda\lambda4041,4824$~\AA{} lines.}
\label{Fig:mgmn}
\end{figure}

The [Mg/Mn] vs [Al/Fe] space has been widely used as a chemical diagnostic to separate the MW in-situ population from the accreted one \citep[][]{Hawkins15,Das20}. This space is shown in Figure~\ref{Fig:mgmn}, where our sample is plotted with a selection of MW halo stars from APOGEE DR17 \citep{APOGEEDR17} in grey and in light blue. In-situ MW stars would occupy the Al-enhanced side of this space, while accreted stars are statistically on average Al-poor. The vertical axis provides a division between evolved and less evolved systems/regions, \ie low-$\alpha$ and  high-$\alpha$, respectively. These four regions in Figure~\ref{Fig:mgmn} are divided by the two black dashed lines. In Figure~\ref{Fig:mgmn}, it is possible to identify the thick disc, \ie the blob in the in-situ high-$\alpha$ region, the thin disc, \ie  the low-$\alpha$ blob, and the accreted systems (low- and high-$\alpha$) as  discussed in \citet{Hawkins15} and \citet{Das20}. Our sample occupies the accreted high-$\alpha$ region, with some stars close to the low-$\alpha$ border.

We want to stress that, although this space has been applied to stars and systems with $\FeH\gtrsim-2.3$ -- the metallicity limits of APOGEE, a datasets that does not contain many stars below $\FeH<-2.0$-- its power to disentangle in-situ from accreted origin might be useful only at metallicities $\FeH\gtrsim-1.5$. We note that $\sim85-90$ percent of the APOGEE stars with $\FeH\lesssim-1.5$ (light blue dots), and reliable measurements of [Mg, Al, Mn/Fe], stands on the "accreted" region of Figure~\ref{Fig:mgmn}. If many VMP stars formed also in the lumpy proto-MW, as proposed by \citet{Belokurov22}, the lack of their presence in the in-situ region of this chemical space would suggest that this diagnostic is inefficient to distinguish between the in-situ and the accreted VMP populations. Alternative chemical diagnostic spaces, \eg replacing Al with Na, as in \citet{Buder22}, would provide an inconclusive answer for similar reasons if applied at the VMP regime. Given the lack of a powerful chemical diagnostic at the VMP regime, we cannot firmly rule out the in-situ scenario solely based  on this chemical space.
Future spectroscopic surveys, \eg WEAVE \citep{WEAVE12,WEAVE24} and 4MOST \citep{4MOST_hrdisk}, will provide a statistically large and homogeneous sample of VMP stars that could  be used  to explore new chemical diagnostics spaces to disentangle the in-situ and the accreted populations.

\section{Hints of chemically peculiar properties}\label{sec:hints}
In this Section, various independent chemical diagnostics are  applied to better understand whether our targets have peculiar chemical properties in comparison with to those of the metal-poor MW halo to known structures. We will show that our kinematical selection present chemical peculiarities, which are not clearly visible from classical [X/Fe] visualisation methods (Figures~\ref{Fig:chems}~to~\ref{Fig:bamg}). However, this peculiar chemical profile is visible when the same kinematical cut is applied to another dataset.

\subsection{Stellar phylogenies and chemo-dynamical outliers}\label{sec:tree_out}

Figures~\ref{Fig:chems}~to~\ref{Fig:bamg} show that the majority of our targets display a very similar chemical abundances. Here, we make use of phylogenetic trees to find chemical outliers, to quantify the level of chemical diversity and evolution in the data, and to reconstruct shared histories if these exist in the dataset \citep{Jofre17, Jackson21,deBrito24}.

Following \citet[and references therein]{walsen24}, we analysed trees built using the distance-based method, which consists of applying the agglomerative clustering algorithm named Neighbour-Joining \citep[NJ]{Saitou1987} on a pairwise Manhattan distance matrix \footnote{The Manhattan distance is the sum of absolute differences between the components of two vectors \citep[see][]{Jofre17}.} as a measure. NJ trees are similar to dendrograms, but the branch lengths are related to the distance matrix. In other words, a long branch implies a large difference between two individuals. NJ trees thus allow us to visualise the hierarchical distribution of a dataset.

To account for the influence of uncertainties on branching analysis, the nodal support, or confidence in each branching point, can be generated through perturbing the chemical abundances with their uncertainties and repeating the NJ analysis. 
From the sampled trees it is  possible to find the Maximum-Clade-Credibility (MCC) tree. This represents the tree that has the highest overall nodal support out of the perturbed abundances and is the most reliable representation of relationships given the data \citep[see][for more discussions]{walsen24}. 

\begin{figure}
\includegraphics[scale=0.28]{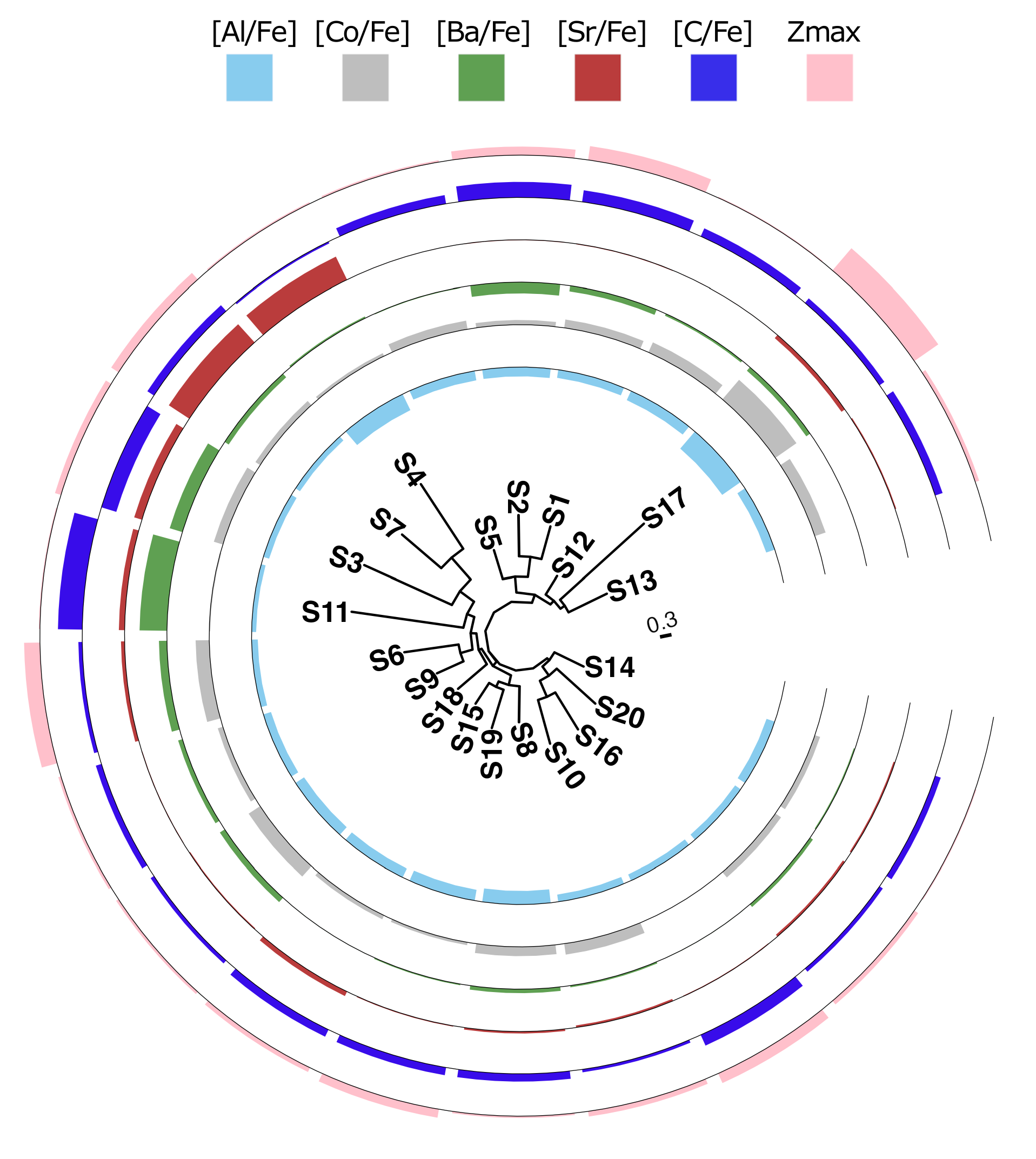} 
\caption{Circular NJ tree built using the distances of 14 chemical abundances of stars in the dataset. Tips represent individual stars (see table online for ID to LAMOST name match). The tree is annotated with abundance ([Al/Fe], [Co/Fe], [Ba/Fe], [Sr/Fe], [C/Fe]) and $\rm{Z_{max}}$ values for each star.}
\label{Fig:tree_circular}
\end{figure}

Figure~\ref{Fig:tree_circular} shows a circular MCC NJ tree built from a Manhattan chemical distance matrix of 14 chemical ratios [X/Fe] and their uncertainties for all the stars in the dataset. 
Coloured bars indicate [Al, Co, Ba, Sr, C/Fe] and $\rm{Z_{max}}$. Four stars significantly stand out, namely J080626.72$+$194212.2 (S4),  J103037.10$+$224124.4 (S7), J225724.46$+$385951.0 (S11), and J110847.18$+$253047.2 (S17). We want to highlight that kinematical properties are not included in the derivation of the branch lengths. 

Two potential outliers, J080626.72$+$194212.2 (S4) and J103037.10$+$224124.4 (S7), are enhanced in Sr ([Sr/Fe]~$\sim1.3$), while having [Ba/Fe]~$\sim0.1$ and $\sim-0.3$, respectively. As already discussed in Section~\ref{sec:dgufd}, the extra production of Sr in these two stars might be explained by their stellar evolutionary phase. Hence, these stars are not outliers in the kinematical properties nor in other [X/Fe] ratios.

The third outlier, J225724.46$+$385951.0 (S11), is a CEMP-s, as shown in Figures~\ref{Fig:carbon}~and~\ref{Fig:chems}. 
Therefore, this star should not be classified as chemical outlier, given the extra-production of Ba and C in a companion.

The fourth star, J110847.18$+$253047.2 (S17), appears to have the largest distance from the other stars based on its large branch. From Figure~\ref{Fig:kine} and from Figure~\ref{Fig:tree_circular}, this star also has distinct kinematics, \ie  $\rm{Z_{max}}\sim10\kpc$ and is not compatible with having a planar orbit. We note that only chemical abundances were considered in the distance matrix. 
In particular, this star stands out for its low content in Al, K, V, Ba, and Eu and for its high content in Co and Zr when compared to the other stars (black circle in Figure~\ref{Fig:chems}). Therefore, J110847.18$+$253047.2 is to be considered the true chemo-dynamical outlier in the sample.
Assuming that J110847.18$+$253047.2 must have had an ancestral chemical (and dynamical) history that differs from the rest, we can use it to root the tree and to analyse the history of the other stars.

\begin{figure}
\includegraphics[width=0.5\textwidth]{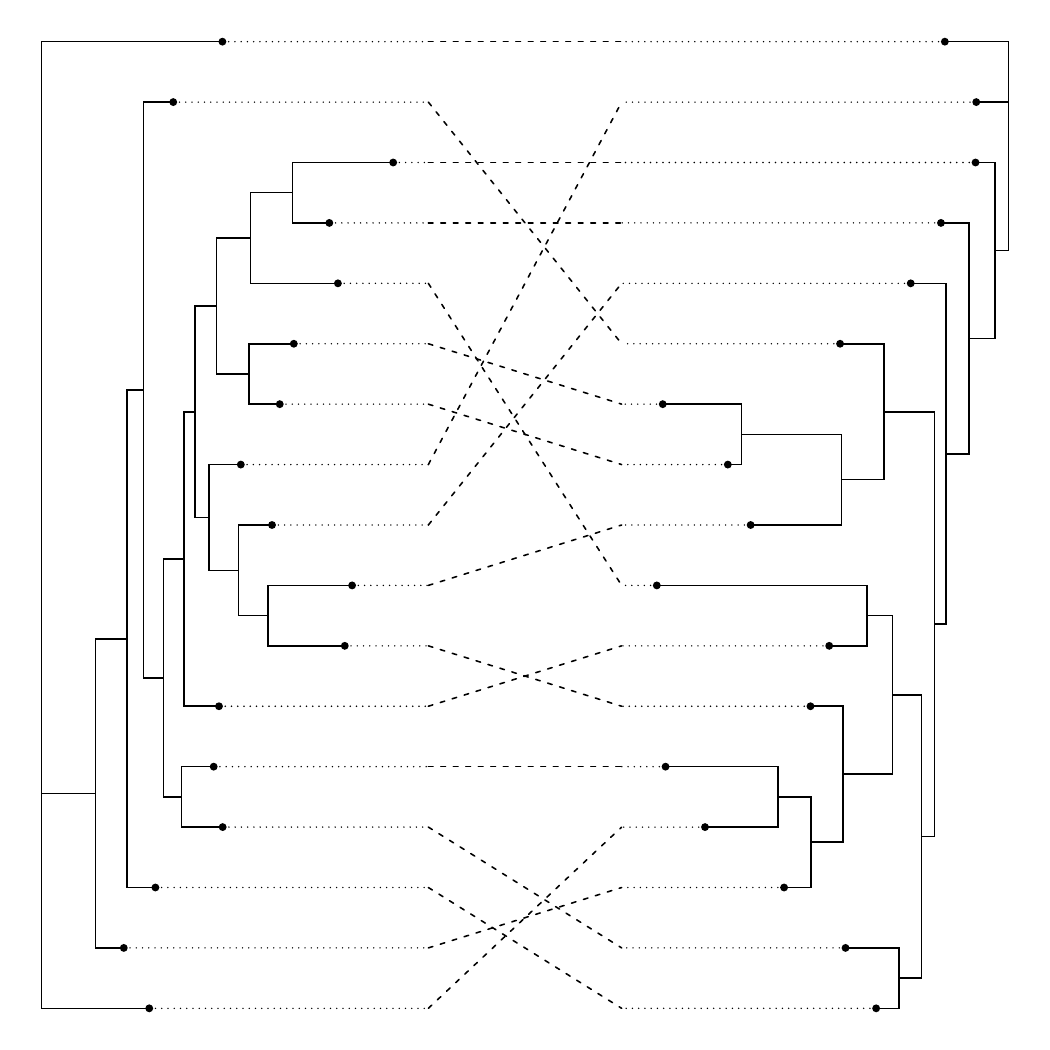}
\caption{Comparison of $\alpha$-capture (left) and Fe-peak (right) tree. Dashed lines connect the same stars in each tree.}
\label{Fig:tree}
\end{figure}

NJ trees do not consider an evolutionary model tailored specifically for astronomical data, and thus should not be used to conclude on a specific history \citep[see discussions in][Tapia-Contreras et al., sub.]{deBrito24,Jofre25}. However, we can use them to gather additional information, like testing whether several types of supernovae contributed to the final [X/Fe] ratios.  We use then NJ trees to test the hypothesis that a tree built only using $\alpha$-capture elements should be different than one from only Fe-peak elements. This might be arising from the fact that supernovae with different energy distribution contributed differently to the final [X/Fe] for $\alpha$- or Fe-peak elements.

To do so, we follow the strategy to compare trees introduced in \citet{deBrito24} and also used in \citet{Jofre25} for the Omega Centauri globular cluster, that considered the Robinson-Foulds (RF) distances. This is a normalised value between 0 (identical) and 1 (totally different) and compares the frequency of nodes that appears in both trees.
Following \citet{deBrito24}, the distance matrix has been perturbed 1000 times to take uncertainties in [X/Fe] into account. The resulting 1000 $\alpha$-capture and Fe-peak NJ trees are then compared to obtain the distribution of the RF distances. 

The mean of the RF distance distribution is $0.91\pm0.04$, indicating that the $\alpha$-capture and Fe-peak NJ trees are significantly different from each other. 
Figure~\ref{Fig:tree} displays the MCC trees built using only [Mg, Si, Ca, Ti/Fe] as input ($\alpha$-capture, left side) and the one built from only [Ni, Cr, Mn, Co/Fe] (Fe-peak, right side). Same leaves/stars are connected through the dashed lines. It is possible to note how these leaves are located at different parts in each tree. 

By calculating RF distances of $\alpha$-capture vs $\alpha$-capture  and of Fe-peak vs Fe-peak trees, we obtain a mean RF distance of $0.69\pm0.11$ and of $0.71\pm0.12$, respectively. This means that, given the uncertainties of our data, the trees can be as similar as $\sim0.7$ and, therefore, a difference of $\sim0.9$ is unlikely to be explained only by uncertainties in the data. This implies that the nucleosynthetic channels imprinted in the chemical distribution of $\alpha$-capture elements is  different  from those driving the diversity of the Fe-peak elements. This is also in line from the \textsc{StarFit} output, \ie hypernovae affect most of the elements up  to the Fe-peak, while high-energy supernovae are mostly imprinted in the lighter ones.

\subsection{Are the chemical abundance dispersions similar to a closed system?}\label{sec:clump}

\begin{figure*}
\includegraphics[width=\textwidth]{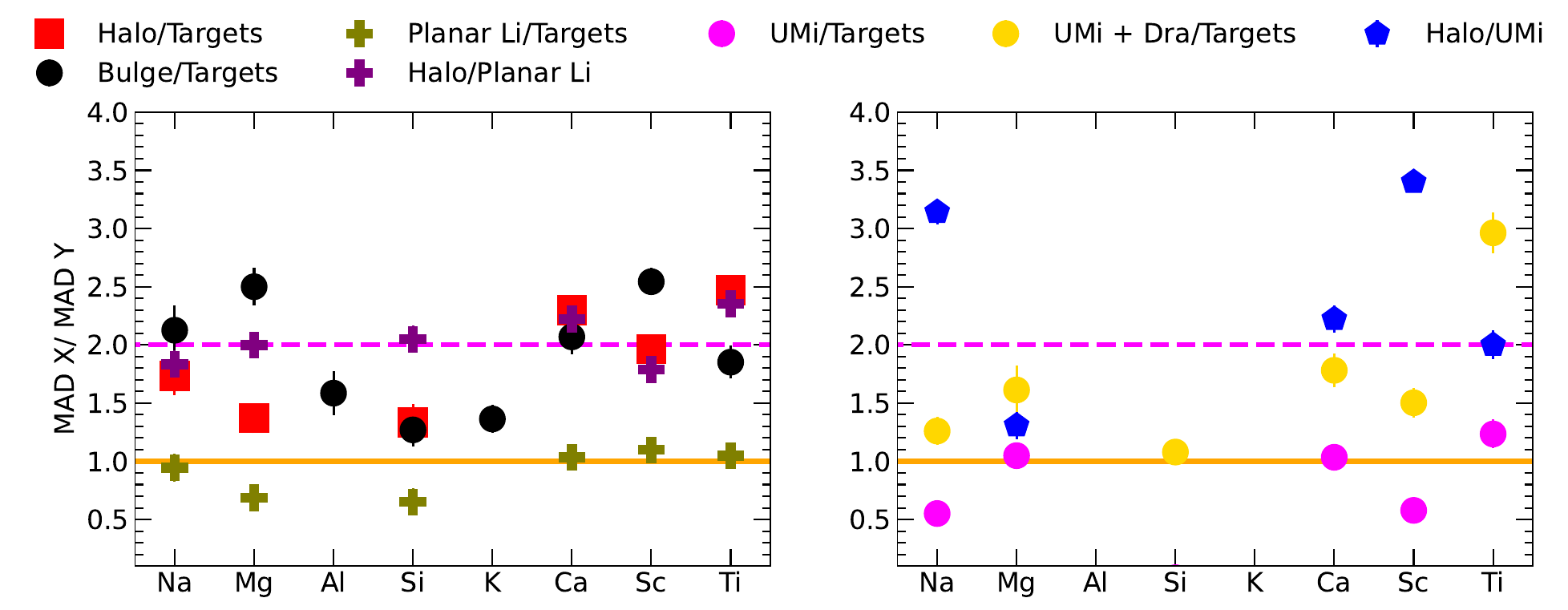}
\caption{Ratios of the Median Absolute Deviation (MAD) as a function of the chemical species.  Stars from the halo \citep{LiH22}, the bulge \citep{Howes14,Howes15,Howes16,Koch16,Reggiani20,Lucey22,Sestito23,Sestito24gh}, the planars, or from dwarf galaxies \citep[from SAGA,][]{Suda08} are selected  from 1D LTE analyses in the optical and to have the same stellar parameters and metallicity range as our targets (no CEMP and no HB). The orange solid line denotes a MAD ratio equal to 1, \ie the same spreads in the two populations; the magenta dashed line marks a MAD ratio equal to 2, \ie the spread in the [X/Fe] of the population in the denominator is half than those of the numerator. Uncertainties on the [X/Fe] have been taken into account when computing the MAD ratios and their uncertainties, which are often smaller than the size of the markers.}
\label{Fig:MAD}
\end{figure*}

The comparison of the chemical abundances shows that the chemical distribution of our targets falls within the region occupied by the compilation of MW halo stars in the classical chemical spaces (Figures~\ref{Fig:chems}~to~\ref{Fig:bamg}). However, MW stars are characterised by a broader chemical dispersion, qualitatively wider than our VMP planar stars.  
It is important to note that, to minimise the spread due to the different assumptions in the chemical analysis between our targets and that of the literature and other systematics between giant and dwarf stars, the halo compilation of Figure~\ref{Fig:chems} has been restricted to stars with similar stellar parameters as our sample ($\log g\leq 3.8$ and $-2.8 \lesssim \FeH \lesssim -1.9$), and measurements from optical high-resolution spectra. 
The relatively small intrinsic dispersions in our targets [X/Fe] ratios is also visible in the phylogenetic trees discussed in  Section~\ref{sec:tree_out}, with the exception for J110847.18$+$253047.2. Here, another independent method is tested to show that our stars have  dispersions in the [X/Fe] smaller than those of the MW halo in the same metallicity regime. In addition, we will show that applying the same kinematical cut to the literature, the new planar sample would possess  smaller [X/Fe] dispersions than the halo.

We first compute the median absolute deviation (MAD) of the [X/Fe] distributions for those elements that, usually, have no trends with metallicity in the VMP regime, \ie the light elements in case Type Ia supernovae are absent \citep[\eg][]{Woosley95,Nomoto13,Kobayashi20}. Other species, such Fe-peak elements might have been produced in some metallicity-dependent ejected supernovae yields, hence showing a trend in their [X/Fe] vs. \FeH{}, which would inflate their MAD (e.g. Mn and Zn as shown in Figure~\ref{Fig:chems}). Successively, we compute the MADs of the halo \citep{LiH22}, of the bulge (see caption for references), of a compilation of planar stars selected from \citet{ZhangMatsuno24} with abundances from \citet{LiH22}, of the Ursa Minor (UMi) dwarf galaxy, and of an ensemble composed of stars from UMi and Draco (Dra). Then, these MADs are divided to the values derived from our targets. The final quantities are called MADs ratios.

These MADs ratios are displayed in Figure~\ref{Fig:MAD},  which has been splitted in two panel for an easier reading. UMi and Dra have been chosen because they are among the systems with the highest number of measured species in the same metallicity range as our sample. The chemo-dynamical outlier, J110847.18$+$253047.2, is removed from this exercise, as well as the other CEMP or Sr-rich stars described in Section~\ref{sec:tree_out}. Peculiar stars have been removed as some of their [X/Fe] might be larger due to binarity (1 CEMP) or evolutionary phases (2 HBs), and not strictly linked to the chemical evolution history. Similarly, HB and CEMP stars have been removed from the comparison compilations. Comparison stars have been selected to reproduce the ranges of \FeH{} and stellar parameters as our restricted sample.

In this MAD ratio space, a value greater than 1 implies a smaller spread in the [X/Fe] distribution for stars in the denominator. Figure~\ref{Fig:MAD} shows that the ratios of the MADs is always larger than $\gtrsim1.4$ when comparing the MW halo from \citet[][red squares]{LiH22} to our targets (red squares). The \citet{LiH22} sample is preferred over the SAGA, since the former is homogeneously observed and analysed, and the instrument has similar resolution as ESPaDOnS. Therefore, the spread observed in the \citet{LiH22} sample should be more similar to the intrinsic dispersion of the MW than what we would obtain from the SAGA dataset. 

The MW halo is a melting pot of several building blocks accreted at early times \citep[\eg][]{Starkenburg17a,ElBadry18,Sestito21}, of later accreted systems \citep[][and reference therein]{Helmi20,Horta23} and of stars formed  in-situ  \citep[\eg][]{Belokurov22}. The lower number of stars in our sample compared with the halo could cause that we are sampling fewer systems than the larger number that comprises the halo, hence ending up with a narrower distribution of chemical abundances. 
In other words, are these results biased by observational limits or low-number statistics? To overcome these additional effects, two Monte Carlo simulations are run. In the first test, a sub-sample of halo stars, composed of the same number of stars and within the same range of metallicities and stellar parameters as our sample, is  randomly drawn for each atomic species for $10^5$ iterations and the exercise of Figure~\ref{Fig:MAD} is repeated. Note that, in this first exercise, the random halo population differs for each elemental species. The probability to find a MAD value similar to or lower than our targets within the halo is generally $<1-2$ percent for each element. The second test quantifies the probability to draw a sub-sample of halo stars to have the same MAD, within an uncertainty of 15 percent, as our sample for all the chemical species. We find that this probability is $\sim0.044$ percent, \ie only 44 draws out of $10^5$ iterations. It is clear that the probability of these stars being part of the general halo distribution and having being picked up by chance is very low. 

Recently, \citet{Lucey22} and \citet{Sestito23}  discussed that the [X/Fe] of the inner Galactic VMP stars are similar to those of the MW halo. Their findings also confirm that multiple formation sites are responsible for the scatter in the [X/Fe], a scenario similarly valid for the MW halo. The MAD ratios in Figure~\ref{Fig:MAD} between the bulge and our targets (black circles)  are always $>1.4$, which indicates  smaller [X/Fe] dispersions for the latter, as the comparison with the MW halo. 

To assess whether the differences in dispersion arise from comparing samples analysed with different methodologies and spectra from different instruments, we identify within the halo sample of  \citet{LiH22} planar stars with the same orbital signature as our targets. Using the kinematical parameters from \citet{ZhangMatsuno24}, we impose an heliocentric distance of $<2\kpc$, an eccentricity from 0.5 to 0.9, and a maximum height from the plane of $3.5\kpc$. The MAD ratios of these planar stars over our targets are shown in Figure~\ref{Fig:MAD} with olive crosses. Na, Ca, Sc, and Ti have comparable MADs among our planars and those from \citet{LiH22}. On the other hand, Mg and Ti show a slightly greater dispersion in our targets. Al and K have not been covered by the observations from \citet{LiH22}.
More impressive is that the MAD ratios values between the MW halo and the new sample of planar stars (purple crosses), both from \citet{LiH22}, are similar or higher than the MW halo vs our targets. In the former comparison (purple crosses) the [X/Fe] ratios in these samples have been measured adopting the same analysis and the same observational setup. In this case, the MAD ratios are $\gtrsim1.8$ for all the species. The spectra of the halo and the planar samples from \citet{LiH22} are qualitatively similar, as they have a SNR~$\gtrsim30$ at  4500~\AA{}, which is high enough to measure  chemical abundances in all the species used in the MAD test. Finding  similar MAD ratios in the \citet{LiH22} samples is an evidence that this kinematical population might also display peculiar chemical properties than the rest of the MW halo.

In a closed system with a well mixed ISM and the absence of Type Ia supernovae, $\alpha-$elements display a small spread and a constant trend with \FeH{} \citep{Kobayashi20}, while the MW halo  should possess a larger spread, given it is composed of various disrupted systems.
The MAD ratios for the MW halo over UMi  are  $>1.3$ for all the species (blue pentagons). In contrast, the MAD ratios of UMi over our targets (magenta circles) oscillate around the value $\sim1$.  
In addition, a compilation of stars from both UMi and Dra is used to check whether the [X/Fe] dispersions in our targets resemble those of two systems with different chemical history. The MAD ratios in this case (gold circles) are $>1$. For Ca and Ti, the ratio is also close or greater than 2. From these comparisons (magenta and gold circles), it  appears that our sample has a chemical dispersion more similar to one system, rather than to those of the sum of two systems with different chemical history.

The previous tests are repeated also dividing our sample into prograde and retrograde, comparing their MADs between them and with the other stellar systems (not shown in Figure~\ref{Fig:MAD}). 
In all the cases, the same result is achieved, \ie our targets show a narrower distribution in the [X/Fe] to those of the MW halo, of the inner Galaxy, and a similar one to those of a closed system. 

\subsection{Highly-eccentric planar stars stand out from the halo in a multi-dimensional chemical space}\label{sec:tsnehalo}

\begin{figure}
\includegraphics[width=0.5\textwidth]{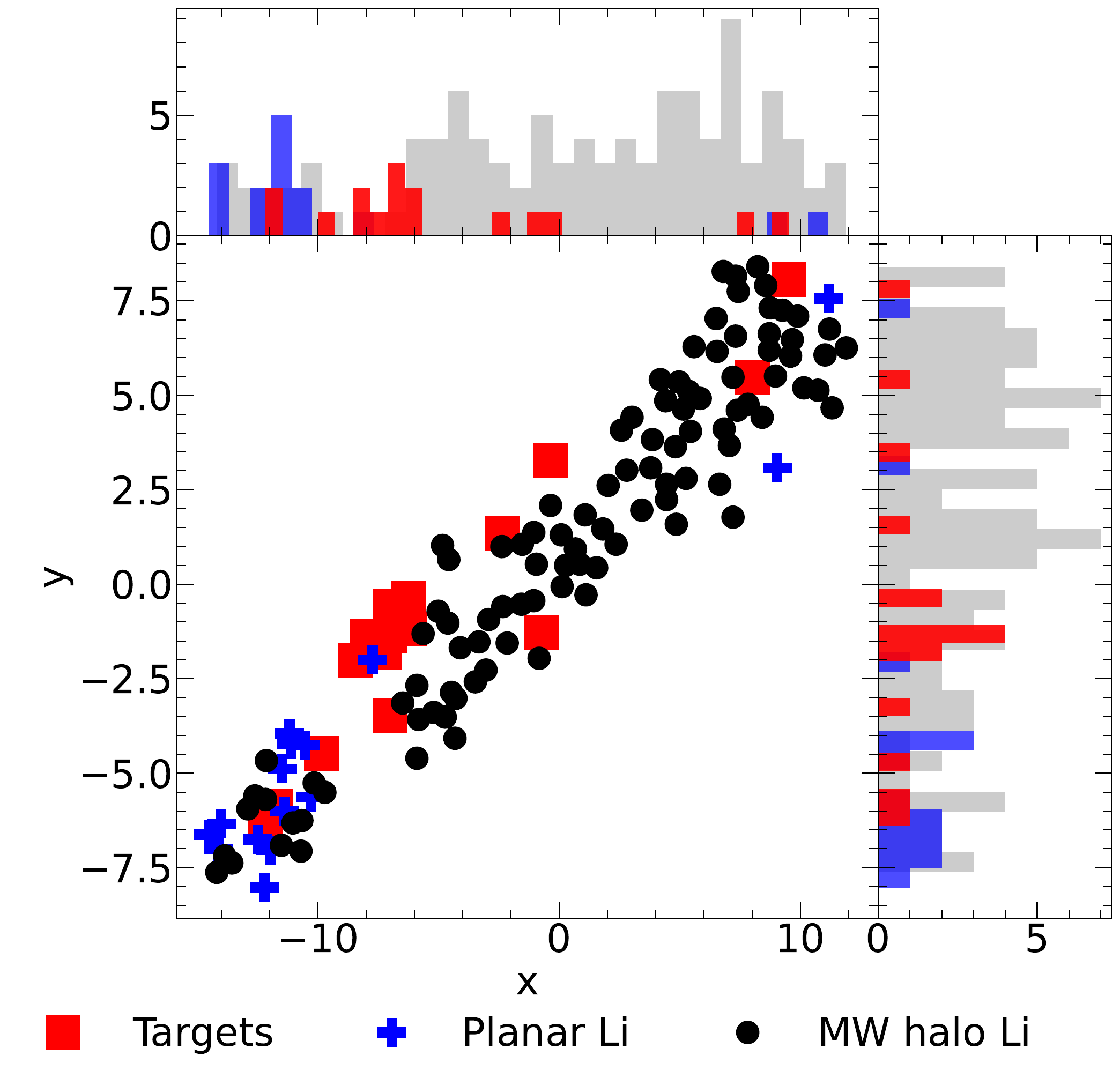}
\caption{t-SNE visualisation chemical space and its histograms. The t-SNE algorithm is used  on a \FeH{} and [Mg, Ca, Ti, Mn, Ba/Fe] chemical space. The chemistry of the halo and planar stars have been analysed by \citet{LiH22}, while their orbital parameters are from \citet{ZhangMatsuno24}. The three samples have the same ranges in metallicities and stellar parameters.}
\label{Fig:tsne}
\end{figure}

To highlight the chemical peculiarity of the planar population, a t-distributed stochastic neighbour embedding (t-SNE) visualisation algorithm is used onto our targets\footnote{This algorithm is preferred over other similar approaches (\eg UMAP), since it focuses on finding local structures.}, the MW halo stars and the planar stars  from \citet{LiH22}. The t-SNE algorithm reduces the high dimensionality of our dataset to 2-dimensional points, grouping those objects that possess similar chemical properties in N-dimensions.  This algorithm uses only chemistry, in particular \FeH{} and [Mg, Ca, Ti, Mn, Ba/Fe], hence no kinematical information. With this exercise we aim to test whether planar stars cluster together and whether they segregate with respect the halo stars.
As the previous exercises, stars with similar stellar parameters and in the same metallicity range have been selected, therefore, reducing systematics due to NLTE effects. As shown in Figure~\ref{Fig:tsne}, planar stars from \citet[][blue crosses]{LiH22} and our sample (red squares) form two clusters. 

The new sample of planar stars from \citet{LiH22} is very well detached from the large majority of MW halo stars \citep[still from][]{LiH22}, also visible in the two histograms. As no kinematical information is used in the t-SNe, this result, obtained with a homogeneously analysed sample, tells us that the high-eccentric planar sample  might have chemical imprints that differ from those of the MW halo. The result using our sample only is not strikingly solid as the previous one, however, the majority of our targets clusters together -- a feature also visible in the histograms --, which might be another indication of the chemical peculiarity of this sample. We remind that our chemical analysis might have different atomic assumptions from \citet{LiH22}, which might not be visible in the classical [X/Fe] space but they can appear in this multi-dimensional space. We retain that such differences might appear as an offset, therefore, affecting the segregations between clusters. 
This offset would not impact previous exercises based  on the dispersion of the chemical abundances.

\subsection{Multi-dimensional chemical comparison with known structures}\label{sec:otherknown}

\begin{figure}
\includegraphics[width=0.5\textwidth]{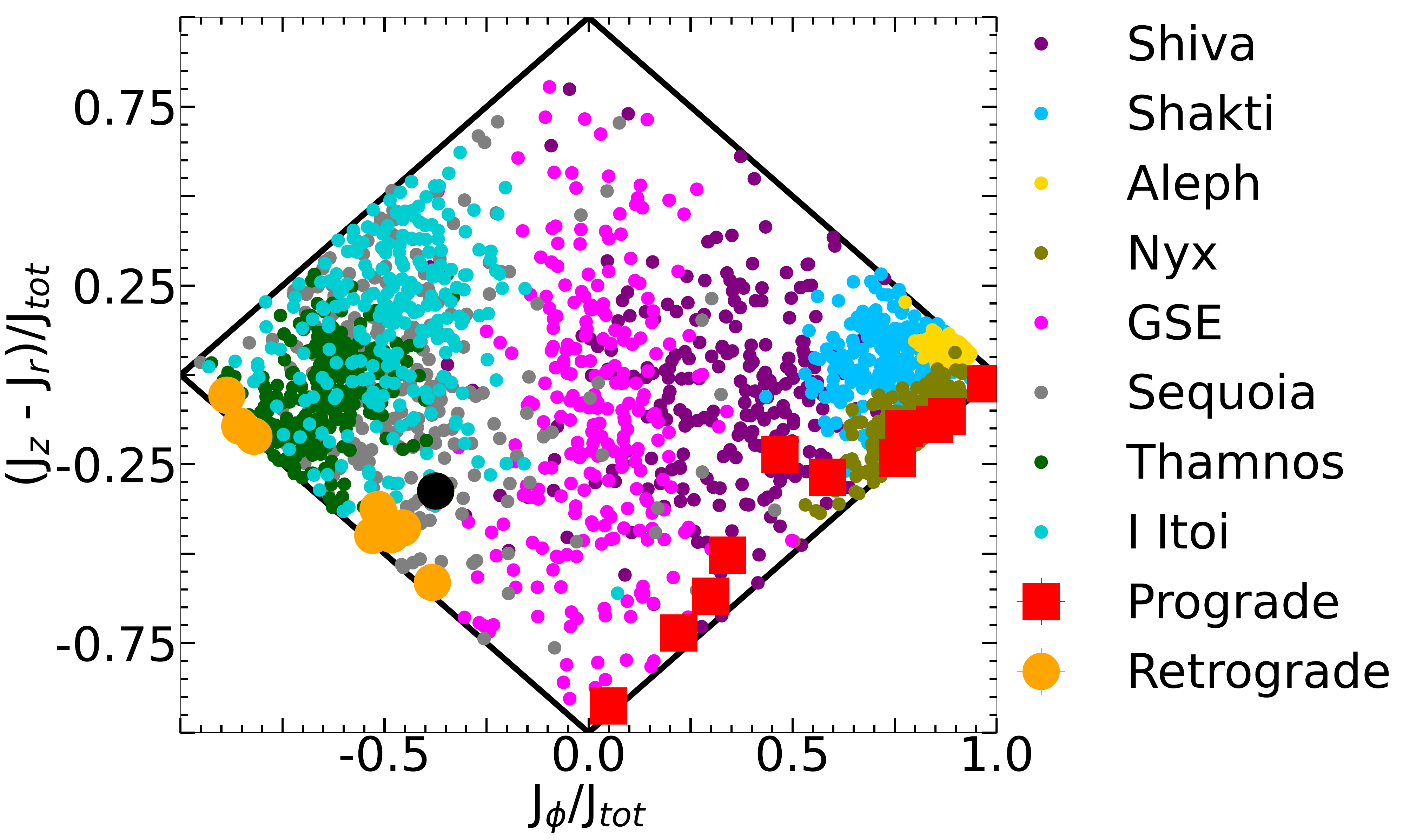}
\caption{The action space. The x-axis is the azimuthal component of the action (equal to L$_z$), while the y-axis is the difference between the vertical and the radial action. Both axes are normalised by the total action, defined as $J_{tot} = J_r +J_z + |J_{\phi}|$. Red squares and orange circles denote our planar stars in prograde and retrograde motion, respectively. The black circle indicates the non-planar J110847.18$+$253047.2. Action values for Shiva and Shakti are from \citet{Malhan24}; Action values for Aleph, Nyx, GSE, Thamnos, Sequoia, and I'Itoi are from \citet{Horta23}. The position of the markers relative to the known structures do not represent real stars as they are randomised from the median of their action quantities.}
\label{Fig:action_accreted}
\end{figure}

Many kinematical structures have been identified since the advent of the {\it Gaia} satellite \citep[][and references therein]{Horta23}. Prograde structures that are  close to the MW plane are Nyx \citep{Necib20}, Aleph \citep{Naidu20}, the Atari disc \citet{Mardini22}, Icarus \citep{ReFiorentin21,ReFiorentin24}, and the recent discovered Shiva and Shakti \citep{Malhan24}. Partially close to the plane , but in retrograde motion, there are Sequoia \citep{Barba19,Matsuno19,Myeong19,Naidu20}, Thamnos \citep{Koppelman19}, Arjuna and I'Itoi \citep{Naidu20}. Heracles \citep{Horta21} and  Gaia-Sausage-Enceladus \citep[GSE,][]{Belokurov18,Helmi18} have an almost null angular momentum and they are linked to the inner Galaxy and the MW halo, respectively, mostly by construction. GSE is  identified as a halo structure, since this is a relatively easier place to look for accreted structures \citep[\eg][]{Carrillo24}, therefore its lower energy part is not well characterised. In the case of Heracles, its stars were selected  imposing a low value for their energy and towards the inner Galaxy. The action quantities for our sample and for some of the known accreted structures are displayed in Figure~\ref{Fig:action_accreted}.

Recently, \citet{Horta23} discussed that the chemical properties of many retrograde structures (\eg Sequoia, Arjuna and I'Itoi) are indistinguishable from those of GSE. Their results bring into question the independence of these substructures, given the fact that these also partially  overlap in the kinematical spaces with GSE \citep[see also results of][separating prograde and retrograde stars and finding results along the same lines]{Kordopatis20}. Therefore, these retrograde structures could be satellites or part of the GSE outskirts. Similarly, \citet{Viswanathan26} proposed that Shakti and Shiva might be the prograde and the lower energy  tail of GSE.

\begin{figure}
\includegraphics[width=0.5\textwidth]{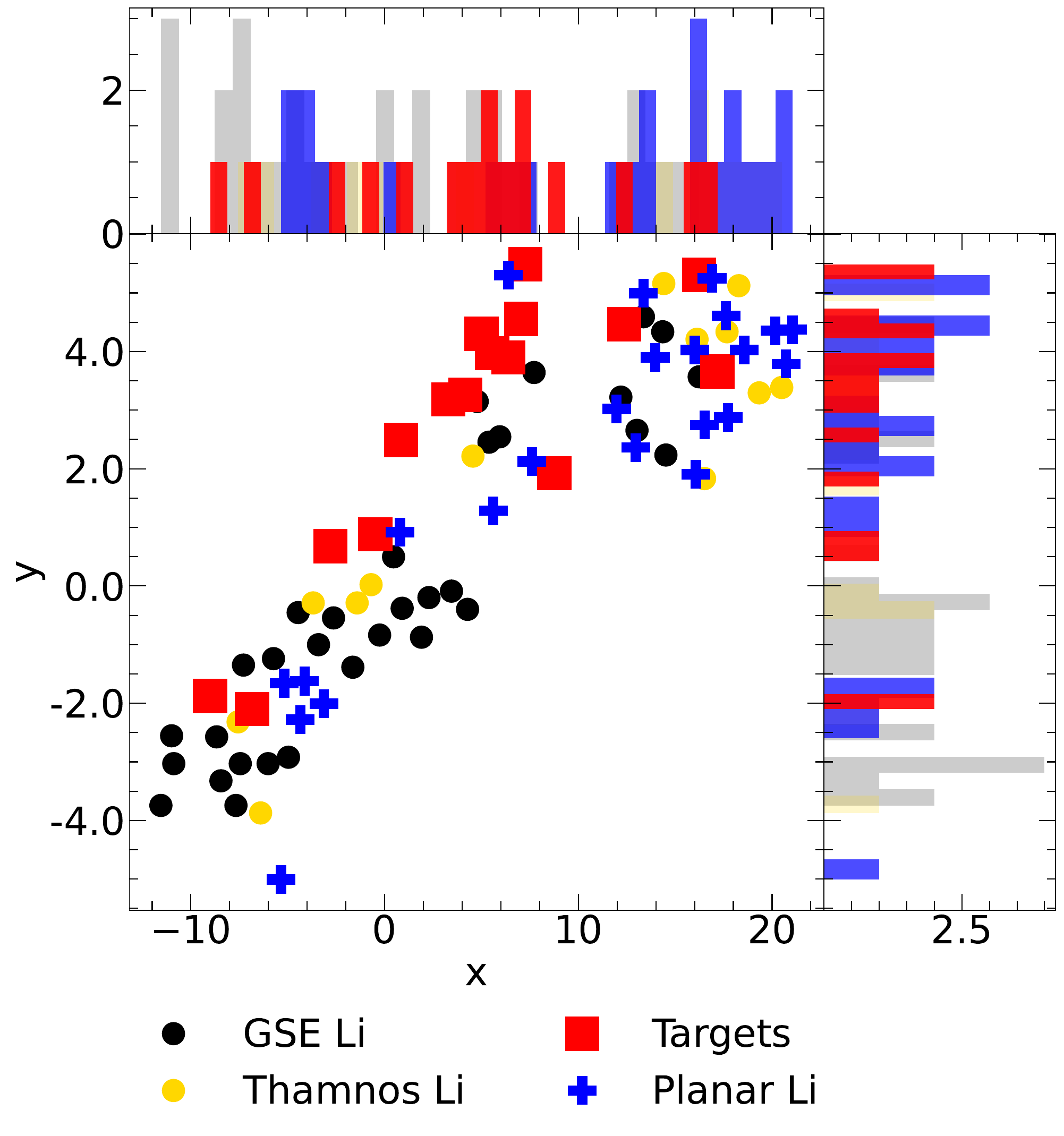}
\caption{t-SNE visualisation multi-dimensional chemical space and its histograms. The t-SNE algorithm is used on a \FeH{} and [Mg, Ca, Ti, Cr, Mn, Ba/Fe] chemical space. The chemistry of stars of GSE, of Thamnos, and of the selected planars have been analysed by \citet{LiH22}, while their orbital parameters are from \citet{ZhangMatsuno24}.}
\label{Fig:tsne2}
\end{figure}

All of these known structures have been chemically characterised mostly using APOGEE, implying NLTE infrared analyses restricted to stars mostly with $\FeH\gtrsim-2.1$. Therefore, a thorough test for possible association between our targets and known structures studied with APOGEE data would be hard to check and beyond the scope of this work. Recently, \citet{ZhangMatsuno24} calculated the kinematical properties for the VMP sample chemically analysed in \citet{LiH22}, dividing it into dynamical groups. \citet{ZhangMatsuno24} characterised the chemical properties of the VMP tail of various  structures, e.g. GSE, Thamnos. As they show and discuss, the [X/Fe] of these structures are in the ranges of those of the MW halo, which is the same result we obtain for our sample. 

In other words, accreted structures might be indistinguishable from the halo at the VMP regime, if a classical comparison of the [X/Fe] vs \FeH{} between systems is adopted, unless the system is enriched or depleted in some of the [X/Fe]. To bypass the limit of a classical chemical comparison, a t-SNE visualisation algorithm, based only on chemistry, is used to compare our sample with planar stars, GSE and Thamnos members kinematically selected from \citet{ZhangMatsuno24} and chemically analysed in \citet{LiH22}. 

This chemical, multi-dimensional space is shown in Figure~\ref{Fig:tsne2}, which also reports the histograms of the two axes. We note that the majority of GSE stars (black circles) have vertical coordinates $<1$ ($\sim66$ percent), while the majority of our targets (red squares) and of the planar sample from \citet[][blue crosses]{LiH22} has vertical coordinates $>1$ ($\sim 75$ and $\sim67$ percent, respectively). The bulk of their distribution separates from GSE. There are, indeed, some outliers that are present in GSE region and vice versa. 
Interestingly, the majority of Thamnos stars ($\sim60$ percent), although less clearly, is located in the same region as the planar ones and separate from GSE. 
Thamnos has been discovered to be strictly retrograde, however, this could be only a selection effect. 
Given the position of Thamnos in the action space (see Figure~\ref{Fig:action_accreted}), \ie its partial overlap our sample, Sequoia and I'Itoi, it might be the case that there are some contamination in the member selection among the various structures.

This t-SNE test might suggest that our sample and the planars from \citet{LiH22}  might have had a different origin from GSE and from Thamnos, if kinematics are also used. However, larger and homogeneous datasets are needed, including observations of both the halo and the disc, to better characterise all the structures with data on the same scale, especially at the VMP regime. This will be possible with forthcoming spectroscopic surveys, such as WEAVE \citep{WEAVE12,WEAVE24} and 4MOST \citep{4MOST_hrdisk}.

\section{A possible origin for these "planar" stars}\label{sec:chemevo}
Our planar sample 
1) has different dynamical properties than the broad distribution of MW halo stars; 
2) presents a narrower chemical distribution than the MW halo, which is visible from the phylogenetic trees (Figure~\ref{Fig:tree_circular}) and from the MAD ratios tests (Figure~\ref{Fig:MAD}); 
3) has a MAD similar to those of a single closed system;
4)  behaves differently than the halo and of some known accreted structures, when considering a multi-dimensional chemical space, clustering together (Figures~\ref{Fig:tsne}~and~\ref{Fig:tsne}); 
5)  shows a steeper slope trend in the [Mn,Zn/Fe] vs [Fe/H] than MW halo star; and 
6) if the same kinematical cut is applied to another sample, the aforementioned chemical properties (vs the halo) are reproduced, highlighting their peculiarities. 

In this section, we will explore whether a single system scenario is supported by cosmological zoom-in simulations, and if it can explain the chemical properties of a larger variety of planar stars, including those in low-eccentric orbits. With galactic chemical evolution models, we will provide an estimate of the baryonic mass in the case one single system is responsible for the formation of our targets and, alternatively, in case two systems are needed to account for the prograde and retrograde sub-samples.

\subsection{The infall of a Galactic building block at early epochs from cosmological simulations}\label{sec:nihao}

\begin{figure}
\includegraphics[width=0.5\textwidth]{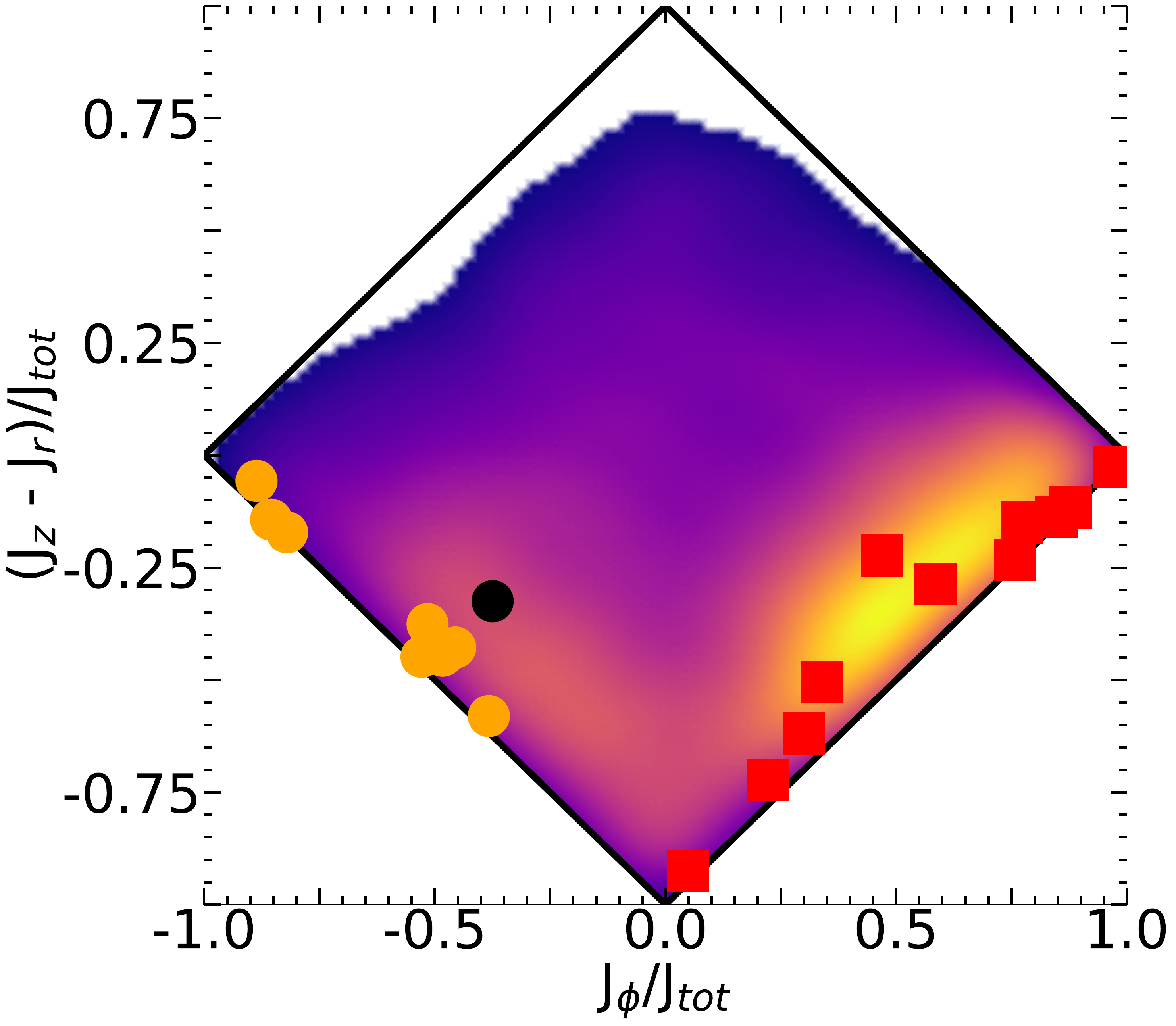}
\caption{The action space similar to Figure~\ref{Fig:action_accreted}. Red squares and orange circles denote our planar stars in prograde and retrograde motion, respectively. The black circle indicates the non-planar J110847.18$+$253047.2. The linear stellar density map of a simulated accreted system in the  MW-like galaxy \textit{g7.55e11} from the NIHAO-UHD simulation suites \citep{Buck20} and studied in \citet{Sestito21}.}
\label{Fig:action_nihao}
\end{figure}

Our sample has stars in both prograde and retrograde orbits. This may raise the question of whether they could have originated from two distinct merger events, if they are of accreted origin. 
The MAD ratios exercise and the phylogenetic trees are in favour of a single-system scenario, unless these stars formed in more systems that possess a very similar or identical chemical enrichment. We look into cosmological zoom-in simulations to understand whether the single-system scenario might be supported or not.

The action space of Figure~\ref{Fig:action_nihao} shows our sample and stellar particles from the NIHAO-UHD simulations \citep{Buck20}, which were used to investigate the origin of the planar VMP stars by \citet{Sestito21}. Specifically, the selected stellar particles belong to a simulated system that was accreted into the forming MW-like galaxy \textit{g7.55e11} at a cosmic time of approximately $\sim3.8$~Gyr.  This simulated accreted system has a stellar mass of $3.2\times10^8\msun$, a gas mass of $6.4\times10^9\msun$, and a dark matter halo mass of $4.9\times10^{10}\msun$, corresponding to approximately 7 (1), 24 (9), and 19 (7) percent of the respective components of the MW-like galaxy at the time of accretion (present time). In terms of the baryonic component, the mass of the accreted system is 21 (7) percent of the simulated MW-like galaxy at the time of accretion (present time). 

The simulations indicate that the accreted system merged with the forming proto-MW-like galaxy with  an injection angle of $<40$ degree. Due to the shallow gravitational potential of the proto-MW-like galaxy and to the lack of a net rotation in the proto-disc, the accreted system was able to disperse its stellar content primarily within the forming proto-disc in a prograde motion, with some dispersal occurring in the retrograde planar and halo configurations. The resulting configuration displays a wide range of angular momenta and eccentricities, similar to those in our sample \citep[the origins of the whole retrograde and prograde planar populations are thoroughly discussed in][]{Sestito21}. 
 The possibility of dispersing stars in both retrograde and prograde orbits is also discussed by the FIRE simulations team \citep{Santistevan21}.
In conclusion, cosmological simulations would allow a single system to disperse its stars in a kinematical configuration similar to our targets.

Based on these simulations, the accretion of such system should have been occurred during the  assembly of the proto-MW, since systems accreted later in the dynamical evolution of a Milky Way-like galaxy (\eg after the formation of the disc) cannot disperse their stars both in a prograde and in a retrograde planar fashion \citep[see][]{Santistevan21,Sestito21}. Interestingly, various theoretical works have shown that dynamical friction and tidal interactions played a significant role in capturing accreted stars into the plane without the need for the merger to be "in-plane" or with a low-injection angle \citep[\eg][]{Penarrubia02,Abadi03,Scannapieco11,Karademir19,Carlberg25}.

\subsection{What about the origin of the other "planar" stars?}\label{sec:compplanar}

Detailed chemical abundance analyses of VMP stars in planar orbits are scarce, and they mostly focus on the prograde and low-eccentricity population, aiming to investigate the primordial tail of the MW disc. Recently, \citet{Dovgal24} re-analysed an EMP star with a quasi-circular prograde planar orbit discovered by \citet{Venn20}, P1836849. \citet{Dovgal24} discuss the kinematics and the  chemical information of P1836849 compared to other 6 quasi-circular prograde VMP planar stars from the literature \citep{Caffau11,Schlaufman18,Sestito19,Cordoni21,Mardini22b}.
Chemical information on these stars is very limited, however \citet{Dovgal24} proposed that a common formation origin is unlikely to be the case, given the diverse chemical patterns across these stars (their Figures~6~and~7), implying  a different chemical evolution. In fact, their sample spans a range of up to $\sim1$ dex in various [X/Fe], a difference well beyond the statistical or systematic error due to a different analysis.

\begin{figure}
\includegraphics[width=0.5\textwidth]{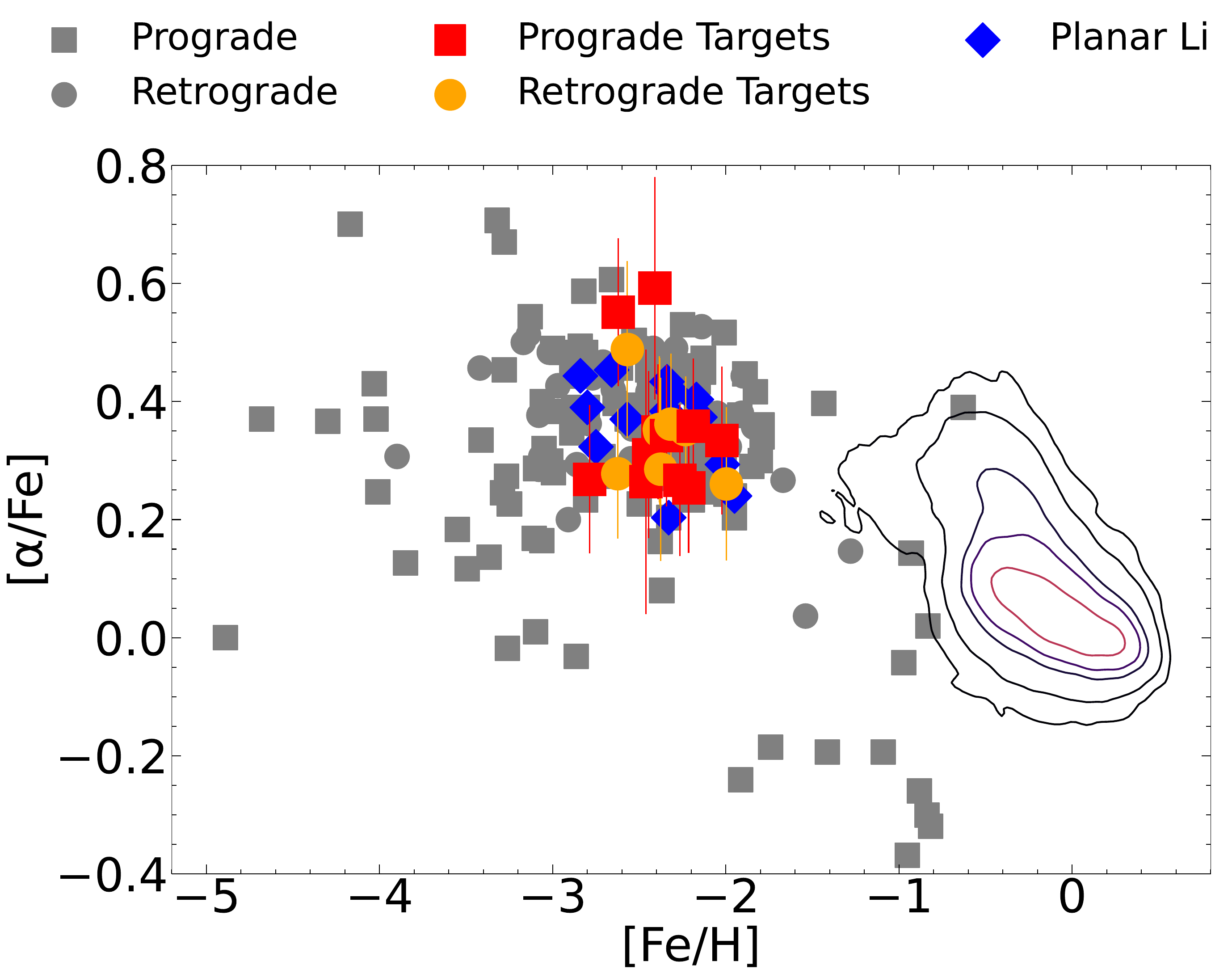}
\caption{[$\alpha$/Fe] vs [Fe/H]. Our prograde and retrograde stars are marked by red squares and orange circles, respectively. Literature planars  with a wide ranges of orbital properties (grey squares and circles) are from \citet{Sestito19}, \citet{Cordoni21},  \citet{Dovgal24}, \citet{FernandezAlvar24}, \citet{Sestito24gh}, and \citet{ZhangMatsuno24}; Blue diamonds mark the sample from \citet{LiH22}, with orbital parameters similar to our sample. Contour lines denote the MW thin and thick discs observed by APOGEE DR17 \citep{APOGEEDR17}; APOGEE stars are selected to have SNR $>70$ and absolute Galactic latitude $<15$ degree.}
\label{Fig:mg_disc}
\end{figure}

We expand the exercise from \citet{Dovgal24}, gathering a larger compilation of planar stars, including objects with prograde and retrograde motion and spanning a wide range of eccentricities. The comparison of their [$\alpha$/Fe]\footnote{[$\alpha$/Fe] is defined as the unweighted mean of [Mg, Ca, Ti/Fe]}  vs \FeH{} is shown in Figure~\ref{Fig:mg_disc}.
 The compilation of planar stars from the literature (in grey squares and circles) includes 8 UMPs \citep[$\FeH\leq-4.0$,][and references therein]{Sestito19}, 21 VMPs from \citet{Cordoni21},  the star from \citet{Dovgal24}, the star from \citet{Sestito24gh}, 127 LAMOST  stars ($\rm{Z_{max}}<3.5\kpc$) analysed in \citet{LiH22} and \citet{ZhangMatsuno24}, and 24 stars with $\FeH\lesssim-0.8$ identified in  \citet{FernandezAlvar24}. Stars with  orbital parameters similar to our sample, analysed in \citet[][chemistry]{LiH22} and in \citet[][kinematics]{ZhangMatsuno24}, are reported with blue diamond markers. Contour lines of the MW thin and thick disc are added to Figure~\ref{Fig:mg_disc}, as a reference level for [$\alpha$/Fe].

The overall [$\alpha$/Fe] distribution does not indicate a clear trend as a function of metallicity, rather a wide spread, \ie $\sim1.2$ dex in $\lesssim4$ dex of metallicity. 
Interestingly, this wide selection of planar stars also displays a large range of values in the other [X/Fe] ratios and we report some examples. The star analysed in \citet{Sestito24gh} has a chemical pattern typical of UFD stars, showing large differences compared to our sample at the same metallicity, \eg of $\sim0.8$, $\sim1.5$, and  $\sim0.5$ dex differences in [Na/Fe], [Ba/Fe], and  [Sr/Fe], respectively. P1836849 \citep{Dovgal24} has a very low  [Na, Mg/Fe] compared to our sample, displaying differences of $\gtrsim0.7$ and $\sim0.3$ dex with the mean of our sample, respectively. The very $\alpha$-poor population, found by \citet{FernandezAlvar24}, has been associated to an accreted origin based on its location in the [Mg/Mn] vs [Al/Fe] space, although it is kinematically similar (by construction) to the MW thin disc \citep{FernandezAlvar21,FernandezAlvar24}. 

The wide ranges of the [X/Fe] among this large compilation of planar stars cannot  be fully reconciled by systematics in the chemical analysis between the various works. To be noted, the chemical abundance analysis in this work, that of \citet{Dovgal24} and of  \citet{Sestito24gh} are performed with the same methodologies and atomic assumptions.
Consequently, we propose that multiple systems with a variety of chemical enrichment histories, from ones more similar to present day UFDs to others similar to classical DGs, contributed to the overall wide group of planar stars. We want to highlight that this compilation includes stars with different kinematical selection criteria, except for a common cut on the maximum height from the plane, which is  $<3.5$~kpc. For this reason, we also test whether the spread in the $\alpha$-elements is still visible using our selected stars with very similar kinematical parameters as our sample, from \citet{LiH22} and \citet{ZhangMatsuno24} (blue diamonds). Using this stricter cut, the bulk of these planar stars are then distributed in [$\alpha$/Fe] between 0.3 to 0.5, hence showing a narrower range than the whole sample and close to the range we obtain for our stars. Another result that might suggest that planar stars with different kinematical properties could have been formed in different systems.

Cosmological zoom-in simulations \citep[\eg][]{Tissera14,Santistevan21,Sestito21} suggest that the VMP broad planar population is composed of various building blocks spanning a wide range of sizes and masses, and likely different chemical evolution. Conversely and if restricting to [$\alpha$/Fe] only, chemo-dynamical simulations from \citet{Khoperskov21} indicate that a wide spread in the [$\alpha$/Fe] distribution among VMP stars (in prograde orbits) can also be a result of the multiphase, complex, and inhomogeneous mixing in the ISM of the forming galaxy. 
Stars in such planar motion can also be moved from their initial configuration, \ie radial migration, by various mechanisms, such as the interaction with the rotating bar \citep[\eg][]{Minchev10}. However, recent modelling of the bar's influence disfavours the possibility that a rotating bar is the main contributor to the planar, and in particular prograde quasi-circular, population \citep{Yuan24,Li_bar24}. 

Metallicities and RVs from {\it Gaia} DR3 \citet{GaiaDR3} have been used to test whether the VMP prograde quasi-circular population is related to the MW thin and thick discs. For example, \citet{ZhangArentsen24} proposed that the VMP prograde population is originated from the superposition of two halo components, one that is stationary (pressure-supported) and one that is rotating at about $\sim80\kms$, which would exclude a common origin with the MW disc. \citet{Bellazzini24} and \citet{Gonzalez24} draw different interpretations using different datasets with kinematics from {\it Gaia} DR3 data, suggesting that this prograde population is instead associated with the thick disc. Both studies discuss that the prograde planar population, even in the VMP regime, is more skewed toward higher angular momenta and is larger in number than the retrograde counterpart, as also shown in previous findings \citep[\eg][]{Sestito20,Cordoni21,Carter21}. \citet{Gonzalez24} conclude that this population would be present even if correcting their kinematics for a rotating prograde halo (by $30-40\kms$), in agreement to what was found in zoom-in cosmological simulations \citep{Sestito21}. Recently, \citet{Hong24} derived photometric metallicities using SkyMapper and Stellar Abundance and Galactic Evolution survey \citep[SAGE,][]{Zheng18} data and confirmed the presence of a VMP prograde quasi-circular population, which they suggest is the remnant of the primordial disc.
However, these previous studies lack  detailed chemical abundance analyses and, therefore, it would be difficult to prove any association of these stars with known structures or with the MW thick disc. A sample of VMP stars in low-eccentric prograde planar orbit has been analysed by \citet{ZhangMatsuno24}. They show that their [X/Fe] are similar to those of the MW halo, although the low [Zn/Fe] and its trend as a function of metallicity can suggest that their progenitor systems might have been small and experienced a slow star formation, giving support to the hypothesis of being part of a particular stellar system with a distinct origin than other halo stars. 

In summary, there are hints that the whole population of planar stars formed in various formation sites, given their different chemo-dynamical properties. However, it is still an open question how to unambiguously clarify any connections between the primordial MW and the various structures at this VMP regime, given the lack of a homogeneous spectroscopic dataset available so far.

\subsection{What if our sample  formed in a single system, a.k.a \textit{Loki}?}\label{sec:snyields}
Given the various tests that indicate the chemical peculiarity of our high-eccentric planar sample and their chemical dispersion similar to a closed system, here, we use galactic chemical evolution (GCE) models to provide some physical and chemical properties in the case our sample is truly formed in a single system, which would be dub as \textit{Loki}\footnote{In the Norse mythology, \textit{Loki} is considered a trickster and  mischievous god/goddess, known for being neither fully good nor evil since his/her main aim was always to create chaos.}, and in case our stars formed in two separate systems, i.e. one for the prograde and one for the retrograde stars.

The GCE model similar to the one used in \citet{FernandezAlvar18} is used  to derive the upper mass ($\rm{M_{up}}$) of the initial mass function (IMF), the integrated yields (Y) from massive stars, the efficiency ($\nu$) of the star formation rate (SFR),  the duration ($\rm{t_{h}}$) of the star formation history (SFH), and the baryonic mass ($\rm{M_{bary}}$). The assumptions to derive these quantities are:
\begin{enumerate}
    \item These stars formed in the same formation site that  initially was made only of pristine gas.
    \item The primordial gas evolves chemically in a closed-box model, \ie no inflows or outflows of gas or stars.
    \item The SFR is proportional to the gaseous mass with  constant efficiency $\nu$, therefore, the SFR decreases with time, as an exponential function.
    \item Semi-instantaneous recycling approximation, \ie massive stars (MS,  $m_i > 9 \msun$) explode as SNe~II immediately after they formed, but the bulk of neutron star mergers (NSM) enrich  the ISM as kilonovae with a delay of about $0.1$~Gyr \citep[\eg][]{Wanajo21}. 
    \item Metallicity-dependent chemical yields for pre-supernovae and Type~II supernovae (SNe~II) are from \citet[][\textit{set F}]{Limongi18}, containing fast-rotating stars up to $150\kms$. Improvements in the n-capture  yields are from \citet{Rizzuti19}.
    \item n-capture elements, like Sr, Y, Zr and Ba, are produced by slow process (s-process, in massive stars) and by rapid process (r-process, in NSM). Their r-process yields are scaled from the Eu yield, considering r-factor by \citet{Simmerer04} who took into account the solar system r-process contribution. 
    \item Eu is produced only through rapid process channel and each NSM ejects $2.5\times 10^{-5} \msun $ \citep{Wanajo14}. 
    \item Most Type~Ia supernovae (SNe~Ia)  explode with a  delay of about 1~Gyr after the formation of their progenitors \citep[see][]{Wanajo21}. However, there is no trace of an $\alpha$-knee in the observed data, therefore the SNe~Ia contribution is null in the chemical evolution of this system.
    \item AGBs stars do not contribute to the enrichment of s-process elements at the very metal-poor regime and within $1$~Gyr from the Big Bang \citep{Doherty14,Karakas16}.
    \item Stars form with the IMF of \citet{Kroupa02} with $\alpha=2.7$ for initial stellar mass $m_i > 1 \msun$.
     \item Our stellar sample between $-3< \FeH <-2$ might correspond to an early portion of a longer star formation history.
     
\end{enumerate}

As the massive metal-poor stars evolve and die, a fraction of their ejected gas is recycled to form new stars. The observed [X/Fe] ratios would be strictly linked to the population of SNe~II and NSM in a given system. Given the [Eu/Fe] ratios in our sample, the total time $\rm{t_{final}}$ needed by the system to form the observed stars should be longer than 0.1Gyr, the necessary time to form NSMs, which are the main source for Eu. As there is no observational evidence for the presence of SNe~Ia in our sample, the $\rm{t_{final}}$ should be shorter than $\sim1.0$~Gyr.

The theoretical yields ejected from SNe~II are integrated  following the adopted IMF, from a mass of $9\msun$ to the upper mass $\rm{M_{up}}$. The upper limit of $\rm{M_{up}}$  is varied until the observed and theoretical [X/Fe] produced by massive stars (\ie no n-capture elements) are matched. We use [Mg/Fe] to constrain the $\rm{M_{up}}$, since Mg and Fe are well determined chemical elements from the observational and theoretical point of view.  
Given all the assumptions in the GCE model, the derived upper mass limit for the SNe~II is $\rm{M_{up}}=55\pm2\msun$.

The upper mass $\rm{M_{up}}$ is strictly linked to the gas mass available to form stars in each burst \citep{Carigi08}, hence to the SFR. 
Using the integrated galactic stellar IMF (index $\alpha=2.7$) as a function of the SFR   \citep[Figure~3 of][]{Weidner13}, the derived SFR is $\sim10^{-1}\msun$~yr$^{-1}$, or $\sim10^8\msun$~Gyr$^{-1}$, which is in agreement with the values predicted for VMP by \citet{Jerabkova18}. 

Since the most metal-rich star formed at $\rm{t_{final}}$ from a gas with $\FeH = -2$, the efficiency of the SFR over time is $\nu= 0.07$ ~Gyr$^{-1}$ \citep[see equation~7 in][]{FernandezAlvar18}. 
The SFR is proportional to $\nu$ and to the initial gas mass, which would lead to an initial baryonic mass (gas and stars)  of $\rm{M_{bary}}\sim 1.4 \times 10^9\msun$. This value is similar to those of classical dwarf galaxies (DGs) in the Local Group, \eg the Small and the Large Magellanic Clouds have a baryonic mass of  $\sim0.9$ and $\sim2.0\times10^9\msun$, respectively \citep[\eg][]{Mcconnachie12}. 
The mass derived by the GCE model is in line with the picture drawn from our observational diagnostics (Figures~\ref{Fig:neutron}~and~\ref{Fig:bamg}), as they suggest that the chemical evolution in these stars was complex and included several s- and r-process sources, very similar to what observed in classical dwarf galaxies, rather than the more pristine ultra-fain dwarfs. To be noted, the baryonic mass is similar to the one possessed by the simulated Galactic building block discussed in Section~\ref{sec:nihao}.

As also discussed in the Section~\ref{sec:hints},  prograde and retrograde stars possess a similar \FeH{} range and  have the same mean [X/Fe], i.e. they share a very similar, if not identical, chemical evolution. If the single system scenario is not truly the case, but rather our sample originated in two systems, one bringing the prograde stars and the other the retrograde ones, what would be their baryonic masses? To address this question, our GCE model is applied to the two sub-samples. Given the same chemistry, the GCE model would suggest that the two systems would have the same properties, i.e. the total  barionic mass would be twice the value of the single-system scenario.

\section{Summary}\label{sec:conclusions}
This work provides, for the very first time, a dedicated detailed chemical abundance analysis of a sample of VMP stars with orbits close to the MW plane. We find that:
\begin{enumerate}[I]
    \item The quality of the ESPaDOnS spectra (Figure~\ref{Fig:spectra}) is very high (SNR up to $\sim220$) and allows us to measure $23$ chemical species (Figure~\ref{Fig:chems}).
    
    \item These stars are relatively close to the Sun and the sample is composed of 11 stars with prograde motion and 9 with retrograde orbits (Figure~\ref{Fig:kine}). One star stands out for its high $\rm{Z_{max}}$, J110847.18$+$253047.2.
    
    \item  The  [X/Fe] ratios of the targets overlap those of the MW halo (Figure~\ref{Fig:chems}).

    \item To reproduce the observed [X/Fe], supernovae yield fitters indicate the presence of neutron star merger events (to account for the n-capture elements), fast-rotating massive stars (for the slow process channels), high-energy supernovae and hypernovae (for the lighter elements) and a lack of SNe~Ia. Observational diagnostics cannot rule out the presence of AGBs as extra source of s-process elements (Figures~\ref{Fig:neutron}~and~\ref{Fig:bamg}).

    \item Observational diagnostics indicate that our sample has  chemical patterns  similar to stars in classical dwarf galaxies rather than those of ultra-faint dwarfs (Figures~\ref{Fig:neutron}~and~\ref{Fig:bamg}), displaying a complex chemical enrichment.

    \item The majority of the low-metallicity stars occupy the accreted region of the [Mn/Mg] vs [Al/Fe] plane (Figure~\ref{Fig:mgmn}). However, VMP stars formed during the earliest phases of the Milky Way’s chemical evolution are also expected to populate this region. At such early times, the chemical enrichment of the proto-Galaxy had not yet produced the abundance patterns that later distinguish in-situ populations, and therefore stars formed in situ may display abundance ratios similar to those typically associated with accreted systems. Consequently, the scarcity of stars in the in-situ region at these metallicities does not necessarily imply an accreted origin. This degeneracy indicates that the [Mn/Mg] vs [Al/Fe] diagnostic becomes inefficient at the lowest metallicities, preventing us from firmly concluding that our sample is of accreted nature.

    \item These targets, with the exception of one star (J110847.18$+$253047.2), show a narrower dispersion in the [X/Fe] than that of the halo and of the bulge at the same \FeH{}. The [X/Fe] dispersions of our targets are very similar to that of a closed system (Figures~\ref{Fig:tree_circular}~--~\ref{Fig:MAD}), and smaller than the case of two formation sites with different chemical enrichment. We find no differences in the chemical abundance dispersions between prograde and retrograde stars. 
    
    \item Looking at similarities in a multi-dimensional chemical space (Figure~\ref{Fig:tsne}~and~\ref{Fig:tsne2}), we see that our planar stars with high eccentricities clump together and well separate from the rest of the halo and from some known accreted structures.

    \item Applying the same kinematical cut to the sample from \citet{LiH22} \citep[see][for orbital parameters]{ZhangMatsuno24}, we obtain the same results as for our sample: high-eccentricity planar stars have a smaller chemical dispersion than the MW halo and they cluster in a multi-dimensional chemical space. This is an important result as \citet{LiH22} contains a homogeneous analysis of stars in the halo,  in known accreted structures, and in planar orbit.

    \item One interpretation of these chemical peculiarities might be that these stars shared a common formation site. Such scenario is allowed by cosmological simulations. The NIHAO-UHD suite of simulations suggests that an infall of a single system into the forming proto-MW could disperse its stars into a wide range of angular momenta, similar to what observed for our targets (Figure~\ref{Fig:action_nihao}). In order to also have stars on retrograde orbits, the accretion events should have happened during the early MW's assembly phase, before the settling of the disc \citep[see][for further details]{Sestito21,Santistevan21}. 

     \item Looking at a wider range of orbital properties among the planar stars (including those with quasi-circular orbits), the [$\alpha$/Fe] vs \FeH{} space (Figure~\ref{Fig:mg_disc}) suggests that multiple systems are needed to account for the wide spread in such chemical ratio. This is also confirmed by the limited availability of other [X/Fe] in the literature. Restricting to eccentricities similar to our sample, the spread in the [$\alpha$/Fe] is reduced and comparable to that of our sample.

    \item If the single system scenario is truly the case for the high-eccentric planar stars (the \textit{Loki} scenario), GCE model would derive a  baryonic mass (gas and stars) of $\sim1.4\times10^9\msun$, a value  similar to those of dwarf galaxies, and also in line with the simulated accreted system from the NIHAO-UHD simulations. Alternatively, if our sample originated in a pair of systems, the simplest case would be one for the prograde and one for the retrograde stars. The pair of systems would share a very similar, if not identical, chemical history and evolution, as suggested by the small MAD and by the GCE model. The total baryonic mass would be twice the case of the single-system scenario.
  
\end{enumerate}

In conclusion, this analysis provides novel results for studies on the Milky Way formation and evolution. We want to emphasise that this analysis has been performed with a  small number of stars, however, this is similar in size to those of other accreted structures at their VMP regime. These planar stars are, undoubtedly, showing hints of a distinct star formation site from normal halo stars that are worthy to be taken into account for further follow-up analysis. It is reassuring that imposing a cut on the orbital parameters in the sample from \citet{LiH22}, we obtain results very similar to those using our sample alone. A fact that is suggesting the peculiarity of this planar population with respect to the MW halo. One intriguing interpretation of the various tests is that these VMPs in planar orbit might be the remnant of an ancient system or, less likely, a pair of systems, sharing a common chemical evolution, that deposited its/their stars during the early Galactic assembly. Certainly, coming large spectroscopic surveys, such as WEAVE and 4MOST, would provide a large and homogeneous dataset that will help to shed light onto the nature and properties of  the various accreted systems and understand if they are truly independent.

\section*{Acknowledgements}
We gratefully acknowledge the anonymous referee for their insightful and constructive comments, which have substantially contributed to the clarification and improvement of several aspects of this manuscript.

The authors wish to recognise and acknowledge the very significant cultural role and reverence that the summit of Maunakea has always had within the Native Hawaiian community. We are very fortunate to have had the opportunity to conduct observations from this mountain. We acknowledge and respect the l\textschwa\textvbaraccent {k}$^{\rm w}$\textschwa\ng{}\textschwa n peoples on whose traditional territory the University of Victoria stands and the Songhees, Esquimalt and WS\'ANE\'C  peoples whose historical relationships with the land continue to this day.

Based on observations obtained with ESPaDOnS at the Canada-France-Hawaii Telescope (CFHT) which is operated by the National Research Council (NRC) of Canada, the Institut National des Sciences de l'Univers of the Centre National de la Recherche Scientifique of France, and the University of Hawaii.

FS thanks Chiaki Kobayashi, Paola Di Matteo, Misha Haywood and Cristina Chiappini for the fruitful feedback received. 

FS thanks the Instituto de Astrof\'isica de Canarias (IAC) for the support provided through its Early Career Visitor program 2023, which enabled a productive research stay and fostered collaborations that were essential for the completion of this project. FS acknowledges funding from the UK Science and Technology Facilities Council through grant ST/Y001443/1. FS and KAV thanks the National Sciences and Engineering Research Council of Canada for funding through the Discovery Grants and CREATE programs. EFA acknowledges support from HORIZON TMA MSCA Postdoctoral Fellowships Project TEMPOS, number 101066193, call HORIZON-MSCA-2021-PF-01, by the European Research Executive Agency. EFA also acknowledges support from the Agencia Estatal de Investigaci\'on del Ministerio de Ciencia e Innovaci\'on (AEI-MCINN) under grants “At the forefront of Galactic Archaeology: evolution of the luminous and dark matter components of the Milky Way and Local Group dwarf galaxies in the {\it Gaia} era” with references PID2020-118778GB-I00/10.13039/501100011033, PID2023-150319NB-C21/10.13039/501100011033 and PID2023-150319NB-C22/10.13039/501100011033. 
LC thanks the support by SECIHTI CBF-2025-I-2048 and UNAM/DGAPA/PAPIIT/IA103326 projects.  PJ, DdBS, CJLE, and SV acknowledge the Millennium Nucleus ERIS (ERIS NCN2021017) and FONDECYT (Regular number 1231057) for the funding. AAA acknowledges support from the Herchel Smith Fellowship at the University of Cambridge and a Fitzwilliam College research fellowship supported by the Isaac Newton Trust. NFM gratefully acknowledges support from the French National Research Agency (ANR) funded project ''Pristine'' (ANR-18-CE31-0017) along with funding from the European Research Council (ERC) under the European Unions Horizon 2020 research and innovation programme (grant agreement No. 834148). ES acknowledges funding through VIDI grant "Pushing Galactic Archaeology to its limits" (with project number VI.Vidi.193.093) which is funded by the Dutch Research Council (NWO). This research has been partially funded from a Spinoza award by NWO (SPI 78-411). PBT acknowledges partial funding by Fondecyt-ANID  1240465/2024 and N\'ucleo Milenio ERIS. TB’s contribution to this project was made possible by funding from the Carl-Zeiss-Stiftung. This research was supported by the International Space Science Institute (ISSI) in Bern, through ISSI International Team project 540 (The Early Milky Way).

This work has made use of data from the European Space Agency (ESA) mission {\it Gaia} (\url{https://www.cosmos.esa.int/gaia}), processed by the {\it Gaia} Data Processing and Analysis Consortium (DPAC, \url{https://www.cosmos.esa.int/web/gaia/dpac/consortium}). Funding for the DPAC has been provided by national institutions, in particular the institutions participating in the {\it Gaia} Multilateral Agreement.
This work has made use of data from the LAMOST survey. Guoshoujing Telescope (the Large Sky Area Multi-Object Fiber Spectroscopic Telescope LAMOST) is a National Major Scientific Project built by the Chinese Academy of Sciences. Funding for the project has been provided by the National Development and Reform Commission. LAMOST is operated and managed by the National Astronomical Observatories, Chinese Academy of Sciences.
This research has made use of the SIMBAD database, operated
at CDS, Strasbourg, France \citep{Wenger00}. This work made
extensive use of TOPCAT \citep{Taylor05}.

\section*{Data Availability}
Raw and reduced spectra are available at the Canadian Astronomy Data Centre (\url{https://www.cadc-ccda.hia-iha.nrc-cnrc.gc.ca}). Tables with chemical abundances, stellar parameters, distances and orbital properties will be available online as supplementary material and at the CDS once the manuscript is published. The galactic chemical evolution model can be shared upon reasonable requests. The present-day snapshot of the NIHAO-UHD cosmological hydrodynamical simulations used here is available at \url{https://tobias-buck.de}.

\section*{Affiliations}
$^{1}$ Centre for Astrophysics Research, Department of Physics, Astronomy and Mathematics, University of Hertfordshire, Hatfield, AL10 9AB, UK\\
$^{2}$ Department of Physics and Astronomy, University of Victoria, PO Box 3055, Victoria BC V8W 3P6, Canada\\
$^{3}$ Instituto de Astrof\'isica de Canarias (IAC), V\'ia L\'actea, E-38200 La Laguna, Tenerife, Spain\\
$^{4}$ Universidad de La Laguna, Dept. Astrof\'isica, E-38200 La Laguna, Tenerife, Spain\\
$^{5}$ Universidad Nacional Aut\'onoma de M\'exico, Instituto de Astronom\'ia, AP 70-264, 04510, Ciudad de México, M\'exico\\
$^{6}$ Instituto de Estudios Astrof\'isicos, Universidad Diego Portales, Av. Ej\'ercito Libertador 441, Santiago, Chile\\
%$^{7}$ Millenium Nucleus ERIS, Chile\\
$^{7}$ Department of Physics and Astronomy, Uppsala University, Box 516, SE-751 20 Uppsala, Sweden\\
$^{8}$ INAF - Osservatorio Astrofisico di Torino, Strada Osservatorio 20, I-10025 Pino Torinese (TO), Italy\\
$^{9}$ Universit\'e C\^ote d'Azur, Observatoire de la C\^ote d'Azur, CNRS, Laboratoire Lagrange, Nice, France\\
$^{10}$ Institute of Astronomy, University of Cambridge, Madingley Road, Cambridge CB3 0HA, UK\\
$^{11}$ Universit\'e de Strasbourg, CNRS, Observatoire astronomique de Strasbourg, UMR 7550, F-67000 Strasbourg, France\\
$^{12}$ Max-Planck-Institut f\"ur Astronomie, K\"onigstuhl 17, D-69117 Heidelberg, Germany\\
$^{13}$ Kapteyn Astronomical Institute, University of Groningen, Landleven 12, 9747 AD Groningen, The Netherlands\\ 
$^{14}$ Instituto de Astrof\'isica, Pontificia Universidad Cat\'olica de Chile, Av. Vicu\~na Mackenna 4860, Santiago, Chile  \\
$^{15}$ Centro de Astro-Ingeniería, Pontificia Universidad Cat\'olica de Chile, Santiago, Chile \\
$^{16}$ Laboratoire d'astrophysique, \'Ecole Polytechnique F\'ed\'erale de Lausanne (EPFL), CH-1290 Versoix, Switzerland\\
$^{17}$ GEPI, Observatoire de Paris, Universit\'e PSL, CNRS, 5 Place Jules Janssen, F-92195 Meudon, France\\ 
$^{18}$ Universit\`a degli Studi di Bologna, via Piero Gobetti 93/2, Bologna, Italy, 40129\\
$^{19}$ Dipartimento di Fisica e Astronomia, Universit\`a degli Studi di Firenze, Via G. Sansone 1, I-50019 Sesto Fiorentino, Italy\\
$^{20}$ Universit\"at Heidelberg, Interdisziplin\"ares Zentrum f\"ur Wissenschaftliches Rechnen, Im Neuenheimer Feld 205,  D-69120 Heidelberg, Germany\\
$^{21}$ Universit\"at Heidelberg, Zentrum f\"ur Astronomie, Institut für Theoretische Astrophysik, Albert-Ueberle-Stra{\ss}e 2, D-69120 Heidelberg, Germany
\bibliographystyle{mnras}
\bibliography{loki}

\appendix

\label{lastpage}

%\bsp	% typesetting comment

\end{document}